\documentclass{aa}  

\usepackage[dvipsnames]{xcolor}
\usepackage{graphicx}
\usepackage{txfonts}
\usepackage{placeins} 
\usepackage{ulem}
\usepackage{wasysym}
\SetSymbolFont{wasy}{bold}{U}{wasy}{m}{n}

\usepackage[
  breaklinks = true,
  colorlinks = true,
  urlcolor   = blue,
  citecolor  = blue,
  linkcolor  = blue,
]{hyperref}
\graphicspath{ {.} }

\newcommand{\Halpha}{{H$\alpha$}}

\newcommand{\FeI}{{\ion{Fe}{I}}}

\newcommand{\FeXII}{{\ion{Fe}{XII} $195$\,\AA}}
\newcommand{\FeXIIb}{{\ion{Fe}{XII} $186$\,\AA}}

\newcommand{\FeXV}{{\ion{Fe}{XV} $284$\,\AA}}

\newcommand{\eparb}{E_\parallel}

\definecolor{cadmiumred}{rgb}{0.89, 0.0, 0.13}

\begin{document}

   \title{Three-dimensional magnetic reconnection in unipolar regions}
   \subtitle{A case study of coronal bright points in the Sun}
   \titlerunning{3D reconnection in unipolar regions}

   \author{Daniel N\'obrega-Siverio\inst{1,2,3,4,5}\corrauth{dnobregasiverio@cea.fr}
          \and
          Fernando Moreno-Insertis\inst{2,3}\email{fmi@iac.es}
          \and         
          Andy S. H. To\inst{6}\email{AndySH.To@esa.int}
          }
          

   \institute{Universit\'e Paris-Saclay, Universit\'e Paris Cit\'e, CEA, CNRS, AIM, 91191, Gif-sur-Yvette, France\\
   \email{dnobregasiverio@cea.fr}
   \and
    Instituto de Astrof\'isica de Canarias, E-38205 La Laguna, Tenerife, Spain
    \and
    Universidad de La Laguna, Dept. Astrof\'isica, E-38206 La  Laguna, Tenerife, Spain
    \and
    Rosseland Centre for Solar Physics, University of Oslo, PO Box 1029 Blindern, 0315 Oslo, Norway
    \and
    Institute of Theoretical Astrophysics, University of Oslo, PO Box 1029 Blindern, 0315 Oslo, Norway
    \and
    European Space Agency (ESA), European Space Research and Technology Centre (ESTEC), Keplerlaan 1, 2201 AZ Noordwijk, The Netherlands
    }

   \date{Received August 28, 2026; accepted October 01, 2026}

\abstract
{Magnetic reconnection is a fundamental mechanism underlying the heating and a wide variety of eruptive phenomena in 
the solar atmosphere; however, characterizing its 3D aspects remains challenging.}
{Our aim is to understand how 3D magnetic reconnection in unipolar regions drives localized heating and plasma ejections.}
{
To this end, we used a self-consistent 3D radiation-magnetohydrodynamic simulation of a coronal bright point (CBP) performed with the 
Bifrost code.
We focused on localized, intense heating events and upflows near a unipolar CBP footpoint associated with its parasitic polarity.
3D reconnection diagnostics were carried out using the parallel electric field and its integral along magnetic field lines.
We then compared the resulting upflows with spectroscopic observations from Hinode/EIS.
}
{
We find recurrent episodes of 3D magnetic reconnection without nulls at the footpoints of loops rooted near the 
CBP’s parasitic polarity.
A detailed analysis of one representative event places the reconnection region at heights $z \in[1.0,2.5]$~Mm,
within a predominantly vertical magnetic field with localized concentrations of electric current.
The dissipation of these currents efficiently heats the plasma to about 2 MK and produces strong gas-pressure gradients 
that drive upflows of a few tens of km~s$^{-1}$.
Blueshifts observed with Hinode/EIS in \FeXV, \FeXII, and \FeXIIb\ near the parasitic polarity of an observed CBP are consistent with this scenario.
}
{The recurrence and efficiency of these reconnection events are essential for heating CBP loops to million-degree 
temperatures.
This form of 3D reconnection may operate more generally in the solar atmosphere, dissipating magnetic stress around strong 
flux concentrations and contributing to heating in apparently unipolar regions.
The resulting pressure-driven upflows offer an indirect spectroscopic diagnostic of localized reconnection heating, 
which future observations with MUSE and Solar-C will be able to test at higher spatial and temporal resolution.
}

\keywords{Sun: photosphere --
         Sun: chromosphere --
         Sun: transition region --
         Sun: corona --
         Methods: observational}
\maketitle
\nolinenumbers 

\section{Introduction}\label{s:introduction}

Magnetic reconnection shapes the dynamics and energetics of the solar
atmosphere by releasing stored magnetic energy into heat, flows, waves, and
accelerated particles.
It is central not only to understanding the coronal heating problem 
\citep[e.g.,][]{Parker:1972,Priest_etal:2002,
Gudiksen_Nordlund:2005a,Klimchuk:2006,Rappazzo_etal:2007,
Parnell_De-Moortel:2012,Judge_Ionson:2024} 
but also to explaining a broad range of solar atmospheric phenomena in which magnetic
connectivity is rapidly reorganized, which sometimes involves magnetic nulls,
separatrices, and quasi-separatrix layers (QSLs).
This includes energetic events such as 
flares \citep[e.g.,][]{Shibata_Magara:2011,Fletcher_etal:2011,Aulanier_etal:2012,Janvier_etal:2013} and
coronal jets \citep[][among others]{Yokoyama_Shibata:1996,
Moreno-Insertis_etal:2008,Pariat_etal:2009,
Moreno-Insertis_Galsgaard:2013,
Archontis_Hood:2013,Wyper_etal:2017,Kumar_etal:2018,Kumar_etal:2019b,JoshiR_etal:2020}, 
as well as smaller-scale events such as Ellerman bombs, UV bursts, and surges
\citep[e.g.,][]{Nobrega-Siverio_etal:2016,Rouppe-van-der-Voort_etal:2017,Danilovic_etal:2017,
Young_etal:2018,
Ortiz_etal:2020,
Hansteen_etal:2019,Peter_etal:2019,Ni_etal:2022}.

Many of these events show a clear photospheric imprint in the form of
magnetic flux cancelation.
Such cancelation is commonly interpreted as a consequence of reconnection
between opposite-polarity flux systems, which is  supported by theoretical and numerical models
\citep[e.g.,][]{Priest_etal:1994,Priest_etal:2018,
Moreno-Insertis_etal:2025,Patsourakos_Archontis:2025,Chen_etal:2026},  and is frequently reported in
observations \citep[][among many others]{Zhang_etal:2001,Green_Wallace:2011,
Reid_etal:2016,Panesar_etal:2016,Panesar_etal:2017,Gosic_etal:2018,Liu_etal:2018,
Tiwari_etal:2019,Kumar_etal:2019a,Bhatnagar_etal:2025b}.
However, magnetic reconnection need not leave a photospheric
signature in the form of local flux cancelation.
In the corona, braided or sheared magnetic fields can develop current concentrations
and undergo reconnection, even when neighboring field lines are
only moderately misaligned.
Such events are particularly subtle  and difficult to isolate 
because they can occur away from magnetic nulls
and moreover may not be related to large changes in magnetic connectivity.
Even so, braided extreme-ultraviolet (EUV) loop structures and small-scale reconnection
outflows in the form of nanojets have been interpreted as possible signatures
of this process
\citep[e.g.,][]{Cirtain_etal:2013,Antolin_etal:2021,
Sukarmadji_etal:2022,Chitta_etal:2022}.
At the same time, numerical models are actively exploring how braiding builds
current layers and intermittent heating
\citep[e.g.,][]{Pontin_Hornig:2015,Reale_etal:2016,Pontin_etal:2017,Bose_etal:2024,
Cozzo_etal:2026a}, as well as how small-angle reconnection can generate nanojet-like
outflows \citep[][]{Cozzo_etal:2025,Sen_Moreno-Insertis:2025,Sen_etal:2026}.
These examples highlight an important challenge: observations can reveal the plasma 
response to fine-scale coronal reconnection, but the underlying changes in 
3D magnetic connectivity cannot be directly observed and can only be quantified in models.

In such fully 3D situations, especially when no obvious null 
or separator marks the reconnection site, diagnosing reconnection requires following the
nonideal electric field along magnetic field lines through
\begin{equation}
    \Xi = \int_L E_\parallel\, dl ,
    \label{eq:xi}
\end{equation}
where \(L\) denotes a magnetic field line \citep[][]{Schindler_etal:1988,Hesse_Schindler:1988,Priest_Forbes:2000,Biskamp:2005}.
A nonzero $\Xi$ indicates that the field line crosses a region where the frozen-field
ideal-magnetohydrodynamic condition is locally broken, and its magnitude can be related
to the magnetic flux transfer through reconnection in that region. 
More precisely, the key information concerning reconnection lies in the spatial variation in $\Xi$ across neighboring
field lines. 
For example, if field lines are traced from a surface (\(S\)) parametrized by coordinates \((x,y)\), then a nonuniform $\Xi(x,y)$ map, that is, one with $\nabla_{x,y}\,\Xi \neq 0$, 
implies a continuous change in the magnetic connectivity of the adjacent field lines across the nonideal
region \citep[][]{Hesse_Birn:1993,Hesse_etal:2005,Wyper_Hesse:2015}.

In the solar atmosphere, coronal bright points (CBPs) provide a natural
laboratory for bridging 3D reconnection diagnostics obtained from theoretical models with plasma
responses that can be targeted by present and future observations.
CBPs are compact coronal structures composed of million-kelvin small-scale
loops, with typical diameters of a few to several tens of megameters and
lifetimes ranging from a few hours to a few days
\citep{Madjarska:2019,Kraus_etal:2023,Kraus_etal:2025}.
They are ubiquitous, with about 570 CBPs detected daily on the solar disk
\citep[e.g.,][]{Golub_etal:1977,Alipour_Safari:2015,Madjarska:2019}.
CBPs are often associated with magnetic flux emergence and the subsequent
evolution of small bipolar or parasitic-polarity regions, with magnetic fluxes
of about $10^{18}$--$10^{20}$~Mx \citep{Mou_etal:2016}.
Because of their ubiquity and significant contribution to 
the high-energy emission of the solar disk, second only to active regions
\citep{Mondal_etal:2023}, CBPs are key objects for understanding how magnetic
energy is released and converted into coronal heating.

In this work, we investigated localized 3D reconnection above photospheric unipolar
areas at the footpoints of the CBP as well as the associated heating and plasma flows.
Since CBPs are compact coronal loop systems, our aim was to better understand how
this process provides a local mechanism for footpoint heating and hot upflows
that may also be relevant to other loop systems in the solar atmosphere.
Section~\ref{s:methods} briefly describes the model used in this work:
a self-consistent 3D radiation-magnetohydrodynamic simulation of a CBP embedded in a coronal hole.
In Sect.~\ref{s:results}, we combine 3D magnetic-connectivity diagnostics
with thermodynamic and force-balance analyses to analyze the selected events
and relate them to 3D reconnection without nulls, localized Joule heating, and
pressure-driven field-aligned upflows.
In Sect.~\ref{s:observations}, we complement the simulation with observations
from the EUV Imaging Spectrometer \citep[EIS;][]{Culhane_etal:2007} on board the 
Hinode satellite \citep{Kosugi_etal:2007}. 
Finally, in Sect.~\ref{s:discussion}, we discuss the implications for
coronal heating via 3D reconnection in apparently unipolar regions
as well as for upflows at the footpoints of coronal loops.

\section{Methods}\label{s:methods}

The 3D CBP numerical experiment, first presented by \cite{Nobrega-Siverio_etal:2023}, was performed employing 
the radiation-magnetohydrodynamics Bifrost code \citep{Gudiksen_etal:2011}.
This code incorporates, in a self-consistent manner, radiative transfer from the photosphere to the corona, together 
with approximate recipes for the primary radiative losses in the chromosphere attributed to neutral hydrogen, 
singly ionized calcium, and singly ionized magnesium \citep[see][]{Skartlien_etal:2000,Hayek_etal:2010,Carlsson_Leenaarts:2012}.
Additionally, it accounts for thermal conduction, optically thin losses, and an equation of state that includes the $16$ 
most relevant atomic elements for the solar atmosphere.
These elements are selected based on either their high abundance or their significance as electron donors at low temperatures.

To obtain a CBP embedded within a coronal hole, we imposed an initially potential fan-spine null-point configuration in a snapshot 
with preexisting developed convection that had reached a statistically stationary equilibrium.
This is our $t=0$.
The parasitic polarity initially has a maximum positive vertical field strength of $142$~G at the solar surface ($z\approx0$~Mm), 
while the coronal hole is mimicked with a magnetic field that becomes asymptotically vertical and uniform in height with strength $-10$~G.
We then allowed this configuration to evolve for 4.5 hours.
The rationale for choosing this magnetic null-point configuration stems from the fact that CBPs are often observed to appear above photospheric 
regions with a parasitic polarity surrounded by a network predominantly of the opposite polarity. 
This scenario typically suggests the presence of a fan-spine magnetic structure with a null point at coronal heights, as inferred from
observations \citep[e.g.,][]{Zhang_etal:2012, Mou_etal:2016, Galsgaard_etal:2017,Madjarska_etal:2021,Cheng_etal:2023,Petrova_etal:2024,Nobrega-Siverio_etal:2025}.
The details regarding the setup of the simulation can be found in our previous publication \citep{Nobrega-Siverio_etal:2023}.

\begin{figure*}[!t]
   \centering
   \includegraphics[width=1\textwidth]{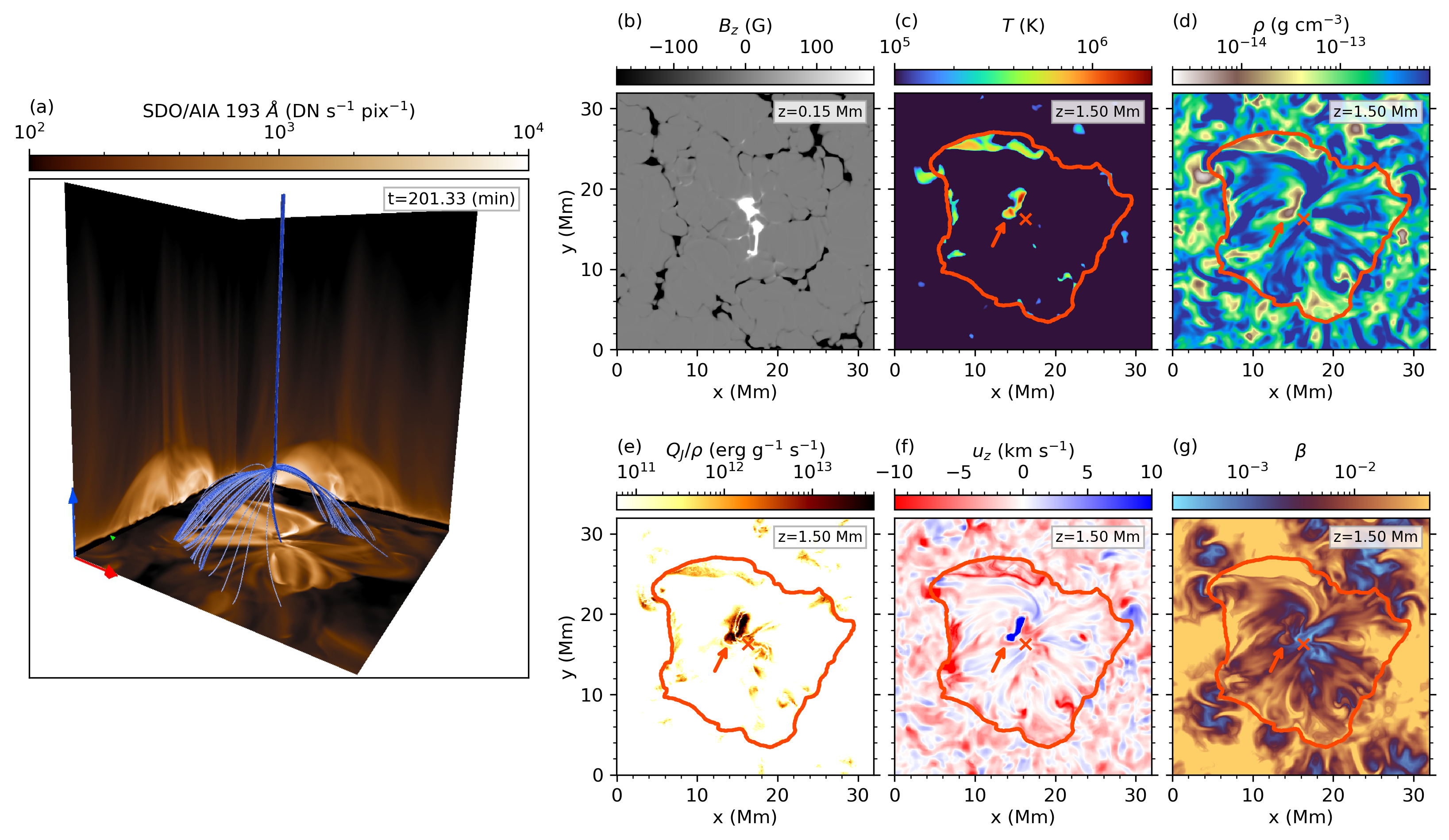}
   \caption{Overview of the CBP simulation at $t=201.33$ min. 
   Panel (a): 3D visualization of the magnetic field lines highlighting the general fan-spine topology of the CBP.
   The red–green–blue axes indicate the $x$-$y$-$z$ directions.
   The panel also includes optically thin synthetic SDO/AIA~193~\AA\ emission integrated along the three Cartesian directions.
   Panel (b): Vertical magnetic field at $z=0.15$~Mm, used as a proxy for the photospheric magnetic field inferred from \FeI\ 6173 \AA\ observations.
   Panels (c)--(g): Horizontal slices at $z=1.5$~Mm showing, respectively, the temperature, $T$; density, $\rho$; Joule heating per unit mass, $Q_J/\rho$; vertical velocity, $u_z$; and plasma $\beta$.
   The arrow indicates the quarter-note-shaped region of interest for our study, while
   the contour and the cross delimit, respectively, the CBP's fan-surface and the location of the inner spine.
   An animation showing the temporal evolution 
   from the initial snapshot ($t=0$~min) up to $t=268$~min is available \href{online}{online}.}
   \label{f:figure_01}
\end{figure*}

\section{Results of the experiment}\label{s:results}
\subsection{Overview}\label{subs:results_overview}
We begin by summarizing the main properties of the 3D numerical experiment, focusing on the aspects that are most relevant 
for the present study. 
In our model, self-consistent stochastic granular motions continuously stress the magnetic field of the CBP, injecting Poynting 
flux into the atmosphere throughout the $4.5$~h evolution \citep{Nobrega-Siverio_etal:2023}. 
Most of the energy deposition within the CBP takes place within the first few megameters above the photosphere. 
There, Joule heating and, to a lesser extent, viscous heating efficiently heat the CBP, especially in the vicinity of the positive parasitic polarity.

Figure~\ref{f:figure_01} provides an overview of the simulation at $t=201.33$ min.
Panel (a) shows a 3D visualization of selected magnetic field lines highlighting the fan-spine topology. 
To provide an observational perspective, the panel also includes background maps of the synthetic Solar Dynamics 
Observatory \citep[SDO;][]{Pesnell_etal:2012}/Atmospheric Imaging Assembly \citep[AIA;][]{Lemen_etal:2012} 193~\AA\ 
emission integrated along the three Cartesian directions \citep[see Appendix C of][for details of how the synthetic emission is calculated]{Nobrega-Siverio_etal:2023}.
The CBP is clearly identified as a system of hot, enhanced-emission loops in this EUV passband.
Panel (b) displays the vertical magnetic field at $z=0.15$~Mm, which is a proxy for what could be inferred from \FeI\ 6173 \AA\ 
observations \citep[see, e.g.,][]{Norton_etal:2006,Moreno-Insertis_etal:2025}, showing that the positive parasitic polarity is located 
approximately at the center of the computational domain.

The remaining panels show horizontal slices at $z=1.5$~Mm.
At this height, panel (c) reveals that the highest temperatures, reaching $T\sim 2$~MK, are concentrated near the center of 
the domain in a distinctive structure with the shape of a quarter note (\quarternote), which is also
apparent in the panels (d)--(f). 
This is the region on which we focused our analysis.
This hot feature is embedded in a rarefied region, as shown in panel (d), where the density is seen to decrease to values 
of about $\rho \sim 5\times10^{-15}$ g cm$^{-3}$, characteristic of coronal plasmas.
Other isolated rarefied regions with relatively high temperatures ($T\lesssim 1$~MK) are also present, primarily in the 
vicinity of the CBP's fan-surface  (see the colored contours).

The Joule heating per unit mass, $Q_J/\rho$, shown in panel (e), exhibits a clear spatial correlation with the hot, rarefied, 
quarter-note-like structure, reflecting the fact that the region is undergoing particularly efficient heating.
For regions in the \quarternote\ structure where $Q_J/\rho>10^{12}$~erg~g$^{-1}$~s$^{-1}$, the Pearson correlation with 
the temperature is 0.7.
The vertical velocity map in panel (f) further shows that this region is dynamically active, with localized upflows 
reaching $u_z \geq 10$ km s$^{-1}$.
The color scale is saturated to ease the identification, but the maximum velocity at this instant is 50 km~s$^{-1}$.
It is also worth noting that the other rarefied regions with relatively high temperatures ($T\lesssim 1$~MK) 
close to the  CBP's fan-surface are predominantly associated with downflows.
Panel (g) contains the plasma-$\beta$ distribution.
At $z=1.5$~Mm, the magnetic field dominates throughout the horizontal domain, with the 
lowest values reaching a few times $10^{-4}$ near the strongest magnetic concentrations.
Within the quarter-note-shaped region, the increased temperature leads to higher values of $\beta$ ($\approx 10^{-2}$), but the plasma remains strongly magnetically dominated.

\begin{figure*}[t]
   \centering
   \includegraphics[width=1.0\textwidth]{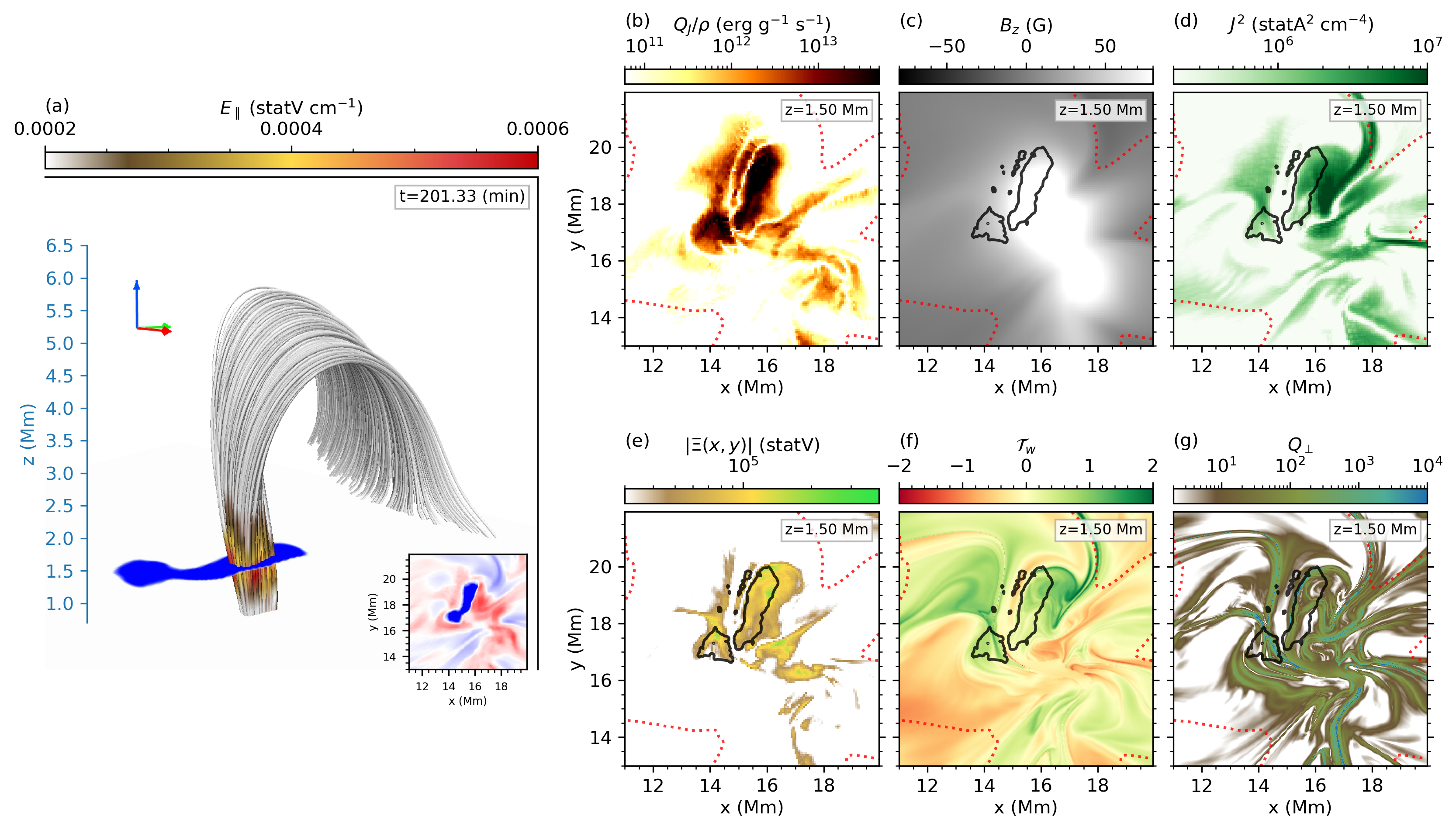}
   \caption{3D reconnection in a unipolar region of the CBP at $t=201.33$ min. 
   Panel (a): Selected bundle of magnetic field lines colored by $\eparb$. 
   The inset contains a zoomed-in $u_z$ map at $z=1.5$~Mm, identifying the quarter-note-shaped region, while the 3D view shows 
   only the strongest upflows ($u_z \geq 10$~km~s$^{-1}$) in blue.
   The red–green–blue axes indicate the $x$-$y$-$z$ directions.
   Panels (b)--(d): Horizontal slices at $z=1.5$~Mm illustrating, respectively,
   $Q_J/\rho$, $B_z$, and the current density squared, $J^2$.
   Panels (e)--(g): $|\Xi|$, $\mathcal{T}w$, and $Q\perp$, respectively, obtained from a field-line tracing with seed points distributed over the horizontal plane at $z=1.5$~Mm with a uniform spacing of $31.25$~km and traced in both directions within $z\in[0.7,6.5]$~Mm.
   In the 2D panels, the solid black and dotted red contours mark $Q_J/\rho=2\times10^{13}$ erg g$^{-1}$ s$^{-1}$ and $B_z=0$, 
   respectively, highlighting that reconnection and heating occur within a unipolar region.
   An \href{online}{online} animation shows the evolution from $t=196.67$~min to $t=201.67$~min.}
   \label{f:figure_02}
\end{figure*}

The simultaneous occurrence of enhanced temperature up to $T\sim 2$~MK, reduced density, strong Joule heating, 
and localized upflows around the parasitic polarity is not unique to the event shown here around $t=201.33$ min. 
Hot, rarefied upflows with similar properties occur repeatedly around the parasitic polarity throughout the CBP evolution, 
with some episodes persisting for several tens of minutes (see the accompanying animation of Fig.~\ref{f:figure_01}).
This suggests that the structure identified here is not a transient peculiarity at the given time, but rather a 
recurrent feature of the CBP. 
Among the events identified in the simulation, 
we focused on the one occurring around $t=201.33$~min (Sects.~\ref{subs:results_reconnection} and \ref{subs:results_upflows}) 
as it provides a representative example of the dynamics and other physical mechanisms common to these episodes.
In Sect.~\ref{subs:results_tracking}, we then broaden the analysis to the full set of identified events, characterizing 
their occurrence and main properties.

\subsection{3D reconnection in a unipolar region}\label{subs:results_reconnection}
We illustrate the physical origin of the hot and rarefied region around the parasitic polarity with the help
of Fig.~\ref{f:figure_02} and the associated animation.
Panel (a) presents a 3D visualization of magnetic field lines colored according to the electric-field 
component parallel to the  magnetic field, $\eparb$, a key quantity for diagnosing 3D magnetic reconnection as mentioned
in the introduction. 
For clarity, here we visualize only field lines seeded within a localized part of the previously identified quarter-note-shaped
upflowing region, namely $x \in [14.3,15.7]$~Mm and $y \in [18.3,20.1]$~Mm. 
A complete perspective of the magnetic connectivity of the full upflow region is provided in Appendix~\ref{app:referee}.
The seed points are placed at $z=1.5$~Mm, and the field lines are traced both upward and downward from these points,
with the downward integration stopped at $z=0.7$~Mm.
The panel also highlights in blue the regions with the strongest vertical upflows ($u_z \geq 10$~km~s$^{-1}$) at $z=1.5$~Mm 
in the vicinity of these lines.
For reference, the $u_z$ map from Fig.~\ref{f:figure_01}f is reproduced 
also here as an inset over the region of interest.
Panel (a) already reveals three important features:
\begin{itemize}
    \item The largest values of $\eparb$ are concentrated near one footpoint of the loop structure, namely in the height range $z\in[1.0,2.5]$~Mm, which encompasses the height $z=1.5$~Mm at which the upflows are detected.
    The animation associated with Fig.~\ref{f:figure_02}, which shows the temporal evolution 
    between $t=196.67$~min and $t=201.67$~min, highlights this fact: the values of $\eparb$ can fluctuate, but they are concentrated 
    around these heights in the same footpoint of the loop structure.
    \item The field lines in that footpoint are relatively concentrated and almost vertical.
    \item The loops reach maximum heights of only about 6~Mm, remaining well below the CBP null point, which is around 
    $z\approx8$~Mm \citep[see][]{Nobrega-Siverio_etal:2023}. 
    This implies that the connectivity changes and associated heating discussed here are unrelated to null-point reconnection, instead pointing to localized 3D reconnection at one loop footpoint (see also Fig.~\ref{f:figure_appendix}).
\end{itemize}

\begin{figure*}[!t]
   \centering
   \includegraphics[width=1\textwidth]{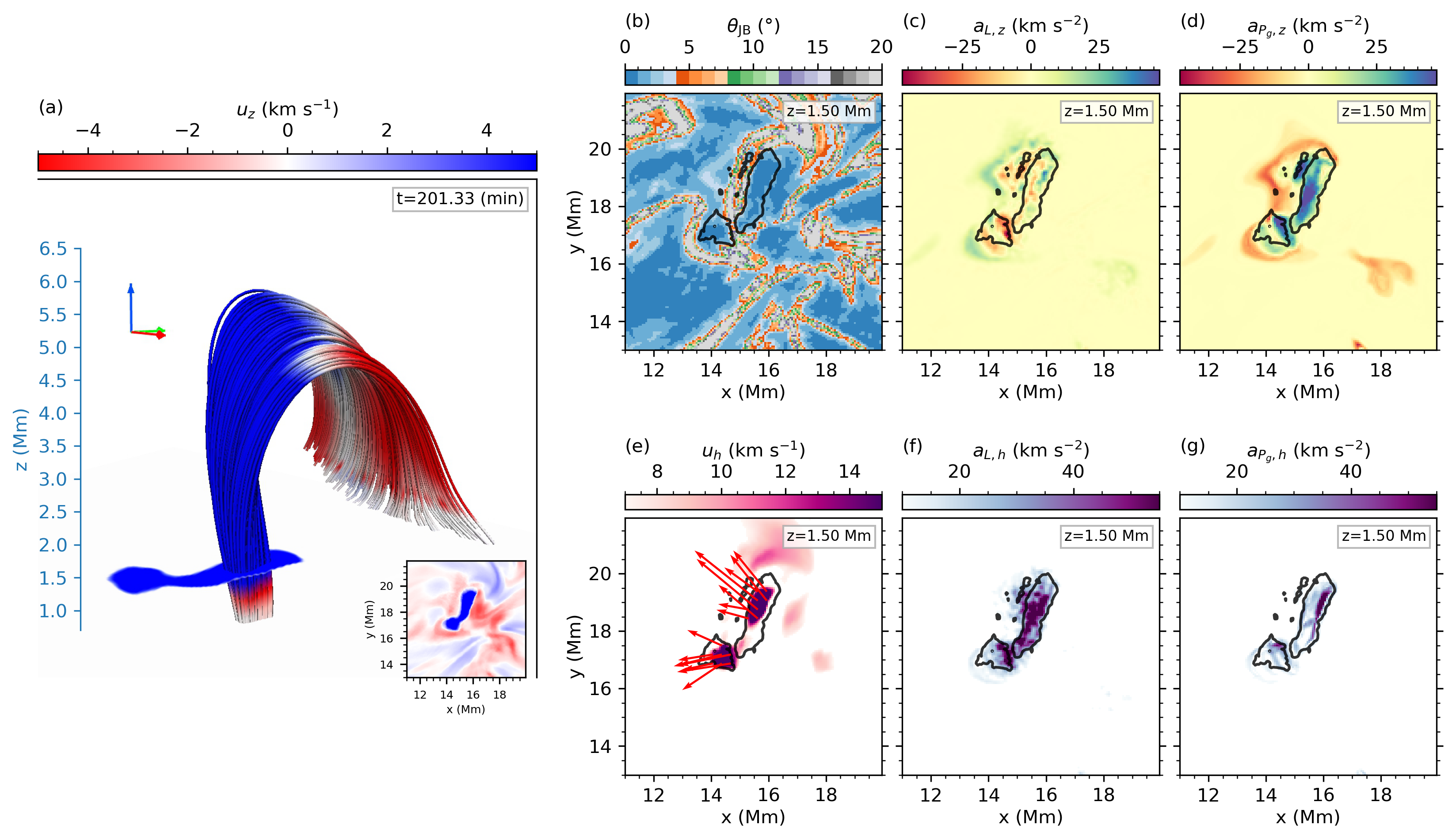}
   \caption{Dynamics of the unipolar region of the CBP at $t=201.33$ min. 
   Panel (a): Same field lines as in Fig. \ref{f:figure_02} but colored by $u_z$, with the scale clipped to reveal weaker downflows.
   Panels (b)--(g): Horizontal slices at $z=1.5$~Mm showing, respectively, 
   the angle between the current density and the magnetic field, $\theta_{JB}$; 
   the vertical Lorentz and gas-pressure-gradient acceleration, $a_{L,z}$ and $a_{P_g,z}$;
    the horizontal velocity ($u_h$), with arrows indicating the direction of the horizontal flows where $u_h>12$~km~s$^{-1}$;
   and
   the horizontal Lorentz and gas-pressure-gradient acceleration, $a_{L,h}$ and $a_{P_g,h}$.
   In the 2D panels, the black line corresponds to the $Q_J/\rho=2\times10^{13}$ erg g$^{-1}$ s$^{-1}$ isocontour.
    An \href{online}{online} animation shows the evolution from $t=196.67$~min to $t=201.67$~min.}
   \label{f:figure_03}
\end{figure*}

The remaining panels of Fig.~\ref{f:figure_02} show horizontal slices at $z=1.5$~Mm zooming into the region of interest.
Panel (b) displays $Q_J/\rho$, which we use as a reference.
Panel (c) contains the vertical magnetic field, $B_z$, and confirms that the efficient heating, outlined by the black contour 
at $Q_J/\rho=2\times10^{13}$ erg g$^{-1}$ s$^{-1}$, occurs within a unipolar area of positive polarity (the red dotted contour marks $B_z=0$).
Panel (d) illustrates that this unipolar region is permeated by electric currents.
However, the strongest heating per unit mass (panel b) does not coincide one-to-one with the largest $J^2$, because it also depends on the local density: weaker currents in rarefied plasma can heat more efficiently than stronger currents in denser regions.

To demonstrate that this is a 3D reconnection event, we performed a new 
field-line tracing over the entire horizontal plane  at $z=1.5$~Mm.
Seed points were uniformly distributed with a spacing of $31.25$~km in both horizontal directions, 
corresponding to half the horizontal grid spacing of the simulation, $\Delta x = 
\Delta y = 62.5$~km.
From each seed point, the field line was traced both upward and downward, 
with the integration restricted to the height range $z\in[0.7,6.5]$~Mm.
This height range was chosen to exclude other nonideal regions, in particular the vicinity of the CBP null point.
Using this set of field lines, we calculated the map of $|\Xi(x,y)|$ (see the introduction), shown in 
panel (e), as well as the field-line twist number, $\mathcal{T}_w$, and the squashing 
factor, $Q_\perp$, shown in panels (f) and (g), respectively (see Appendix~\ref{app:useful_quantities} for the 
definition of the latter two quantities).
Panel (e) shows that the region of enhanced $Q_J/\rho$ is co-spatial with a region of large and non-uniformly distributed values of $|\Xi(x,y)|$, 
supporting the interpretation that the heating is associated with 3D magnetic reconnection.
In fact, our $|\Xi(x,y)|$ distribution is reminiscent of the $\Xi$ maps discussed by \cite{Wyper_Hesse:2015} 
for spatially inhomogeneous $\eparb$, in the context of their theoretical analysis of fragmented current layers.
Within the region of interest, $|\Xi(x,y)|$ reaches values on the order of $10^5$ statV, corresponding to a magnetic flux transfer rate of approximately 
$3\times10^{15}$~Mx~s$^{-1}$ (see the discussion about these values in Sect.~\ref{s:discussion}).

The values of $\mathcal{T}_w$ within the quarter-note-shaped region, shown in panel (f), remain mostly within
$\mathcal{T}_w\lesssim 1$, confirming what is suggested by 
the 3D visualization in panel (a): the magnetic field structure is barely twisted.
The squashing factor, $Q_\perp$, in panel (g) reveals a highly structured distribution of QSLs in this part of the CBP.
In particular, large values of $Q_\perp$ appear in thin, filamentary structures that intersect and surround the quarter-note-shaped 
region, indicating strong gradients in field-line mapping between neighboring field lines (see the discussion in  Sect.~\ref{s:discussion}).

\subsection{Reconnection-driven pressure imbalance flows}\label{subs:results_upflows}
Once we demonstrate that the heating in the region of interest is associated with the dissipation of currents 
through purely 3D reconnection, we will show that the upflows are a consequence of this heating and are primarily 
driven by gas-pressure gradients.
Figure~\ref{f:figure_03}a contains a 3D  visualization of the same magnetic field lines as in the 
previous figure, now colored by the vertical velocity $u_z$.
The color scale is intentionally clipped between $\pm5$ km s$^{-1}$ so that the weaker downflows remain 
visible despite the much stronger upflows.
The field lines reveal a coherent flow pattern in which plasma is accelerated upward from the central part 
of the CBP and subsequently drains downward toward the outer footpoints, close to the CBP's fan-surface. 
A second region of weaker downflows is also present immediately beneath the localized heating site.

\begin{figure*}[!t]
   \centering
   \includegraphics[width=1\textwidth]{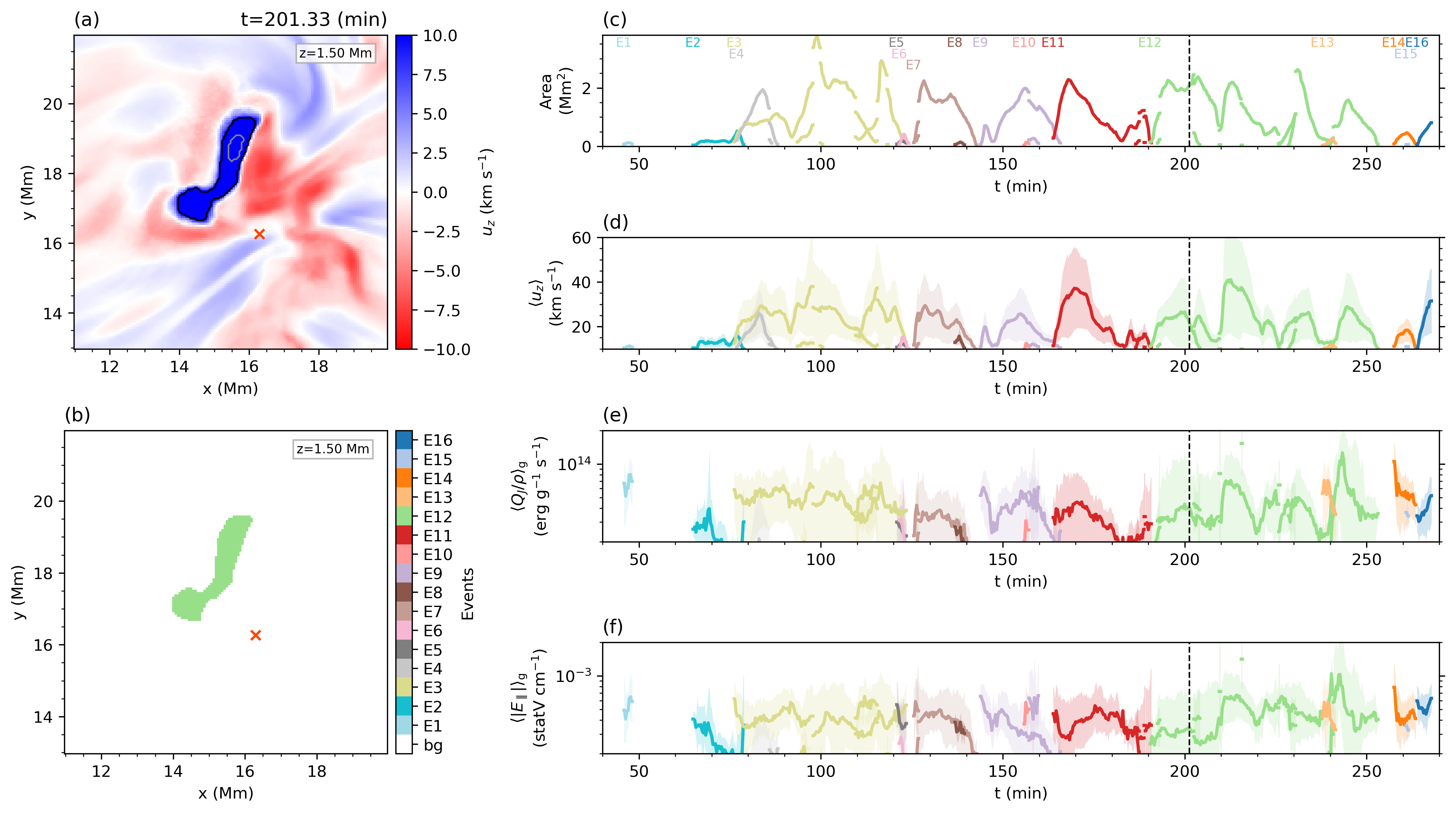}
   \caption{Properties of the upflows identified at $z=1.5$~Mm.
   Panel (a): Vertical velocity, $u_z$, at $t=201.33$~min, with contours delimiting $u_z=10$~km~s$^{-1}$ (black) and
   $u_z=40$~km~s$^{-1}$ (gray).
   Panel (b): Detected upflow colored according to the event to which it belongs.
   In both panels, the location of the inner spine is marked with a cross.
   Panels (c)--(f): Temporal evolution of 
   the area of the upflows;
   their mean vertical velocity, $\langle u_z\rangle$; 
   their geometric mean Joule heating per unit mass, $\langle Q_J/\rho\rangle_{\mathrm{g}}$; 
   and their geometric mean absolute parallel electric field, $\langle|E_\parallel|\rangle_{\mathrm{g}}$, respectively.
   The shaded regions indicate one standard deviation.
   The vertical dashed line marks the time shown in panels (a) and (b).
   An animation showing the evolution 
   from $t=40$~min up to $t=268$~min is available \href{online}{online}.}
   \label{f:figure_04}
\end{figure*}

The remaining panels of Fig.~\ref{f:figure_03} show horizontal slices at $z=1.5$~Mm for various complementary quantities.
Panel (b) presents the angle between the current density and the magnetic field, that is,
\begin{equation}
\theta_{JB} =
\arccos \left(
\frac{|\vec{J}\cdot\vec{B}|}{\left|\left|\vec{J}\right|\right|\,\left|\left|\vec{B}\right|\right|}
\right),
\end{equation}
so $\theta_{JB}=0^\circ$ corresponds to a locally force-free configuration.
In the region of interest, the angle remains very small, typically $\theta_{JB} \lesssim 1^\circ$, indicating 
that the magnetic field is close to a force-free state.
However, this does not imply that the Lorentz force is necessarily negligible compared with the gas-pressure-gradient 
force, particularly in the low-$\beta$ environment considered here (see Fig.~\ref{f:figure_01}g). 
Because the Lorentz force is perpendicular to the magnetic field, which in this region is predominantly vertical, 
the Lorentz force is expected to act mainly in the horizontal plane.
To examine the force balance associated with the vertical flows, in panels (c) and (d), we compare the vertical 
acceleration produced by the Lorentz force, 
\begin{equation}
    a_{L,z} = \frac{(\vec{J}\times\vec{B})_z}{\rho},
\end{equation}
and that caused by the gas-pressure gradient, 
\begin{equation}
    a_{P_g,z} = \frac{-(\vec{\nabla} P_g)_z}{\rho},
\end{equation}
respectively.
In the region of interest, the vertical Lorentz-force contribution is weak. 
In contrast, the gas-pressure-gradient term displays strong positive values that far exceed the downward solar 
gravitational acceleration ($g_{\odot}=-0.274$~km s$^{-2}$), implying a net upward acceleration at this height. 
These results indicate that the upflows are primarily driven by the vertical gas-pressure gradient associated 
with the localized heating produced by 3D magnetic reconnection.
The localized pressure enhancement also drives weaker secondary downflows immediately beneath the 
heating site (see panel a), where the gas-pressure gradient points downward.

Turning now to  the horizontal flows, we encounter a different balance of forces.
Panel (e) shows the horizontal velocity, $u_h$, with superimposed arrows to indicate the direction of 
the horizontal flows where $u_h>12$ km s$^{-1}$.
To identify the mechanism responsible for these motions, panels (f) and (g) contain the magnitudes 
of the horizontal acceleration associated with the Lorentz force and with the gas-pressure gradient, respectively,
\begin{equation}
a_{L,h} = \frac{\left|\left|\left(\vec{J}\times\vec{B}\right)_h\right|\right|}{\rho},
\end{equation}
and
\begin{equation}
a_{P_g,h} = \frac{\left|\left|(\vec{\nabla} P_g)_h\right|\right|}{\rho}.
\end{equation}
In the region of interest, $a_{L,h}$ is generally much larger than $a_{P_g,h}$, and its spatial 
distribution evolves rapidly in time (see the accompanying animation).
In short, the predominantly vertical magnetic field naturally separates the dynamics: the gas-pressure 
gradient drives the vertical upflows, while the Lorentz force dominates the horizontal motions.

\subsection{Recurrence and properties of the upflow events}\label{subs:results_tracking}
The previous sections establish the physical sequence for a representative episode of one of the hot and rarefied upflows: 
localized 3D reconnection, efficient Joule heating, the development of a pressure imbalance, and the subsequent 
acceleration of plasma along the magnetic field. 
We next broadened the analysis to the full time series to determine how recurrent these upflows are and 
to characterize their main properties.
For this purpose, we identified connected upflow patches at $z=1.5$~Mm containing at least 10 adjacent pixels with $u_z>10$~km~s$^{-1}$.
We linked these patches temporally and grouped them into individual upflow events (see Appendix~\ref{app:tracking_method} for details).
Figure~\ref{f:figure_04} and the associated animation show the resulting events and their temporal evolution.
Panels (a) and (b) contain the vertical velocity and the corresponding patch detection.
During approximately the first 45~min, the atmosphere is still adjusting to the imposed fan--spine magnetic configuration.
Some upflows are already present during this initial phase, but they do not simultaneously satisfy 
our adopted selection criteria of $u_z>10$~km~s$^{-1}$ and $\geq 10$ connected pixels.
The first identified significant upflow event appears at $t=46$~min, with a mean vertical velocity just above 10~km~s$^{-1}$, 
after which upflow events recur throughout most of the analyzed interval.
The algorithm identifies 16 distinct upflow events, some of which undergo splitting 
and merging during their evolution.
The position of the inner spine is also indicated in panels (a) and (b) and in the accompanying animation.
The detected upflows generally remain spatially separated from it, although occasional overlap occurs, for instance around $t\simeq165$~min during event E11.

With respect to the size, as shown in panel (c), the instantaneous areas of the detected islands vary considerably during the evolution, reaching sizes
up to $\approx 3$~Mm$^2$.
The lifetimes of the upflow events range from only $\approx1$--$2$~min 
(e.g., events E15 and E10) to as long as $62$~min 
(e.g., E12, to which the upflow event analyzed in this paper belongs).
For reference, the loops associated with these upflows have typical lengths of about $13$~Mm.
Representative values in the upflow region at $z=1.5$~Mm, $B\sim50$--$80$~G (see Fig.~\ref{f:figure_02}c) and
$\rho\sim0.5$--$2.0\times10^{-14}$~g~cm$^{-3}$ (see Fig.~\ref{f:figure_01}d), correspond to Alfvén speeds of order
$1000$--$3200$~km~s$^{-1}$.
The resulting Alfvén travel time is therefore only of order
$4$--$13$~s, much shorter than the typical upflow lifetimes identified.

Concerning the mean vertical velocities of the upflow events (panel d), they are typically a few tens of km~s$^{-1}$, 
while substantially higher velocities, up to approximately $60$--$80$~km~s$^{-1}$, 
can occur locally within individual islands (see, e.g., events E3, E11, and E12).
Interestingly, the larger islands tend to be associated with higher mean upflow velocities, with a Pearson correlation coefficient
$r=0.71$.
Panels (e) and (f) contain the evolution of the geometric mean for the Joule heating per unit mass,
$\langle Q_J/\rho\rangle_g$, and absolute parallel electric field,
$\langle|E_\parallel|\rangle_g$, respectively.
These two quantities are strongly correlated, with a Pearson correlation coefficient of $r=0.89$.
Instead, the correlation between the velocity and heating curves is weaker, with $r=0.50$.
This behavior seems to be consistent with the nonideal electric field $\eparb$ and Joule dissipation being closely associated 
with the local current structures, while the upflow represents the subsequent dynamical response, 
mediated via the gas pressure gradient, of the plasma to the energy deposition (see Sect.~\ref{subs:results_upflows}).

\section{Results from observations}\label{s:observations}
Having analyzed the 3D reconnection, associated heating, and resulting plasma flows, we next investigated whether spectroscopic signatures consistent with the simulated upflows can be identified in observations of CBPs. 
Such observations not only provide an opportunity to assess the validity of our numerical results but can also establish an observational basis for future high-resolution studies.

\begin{figure*}[!t]
   \centering
   \includegraphics[width=1\textwidth]{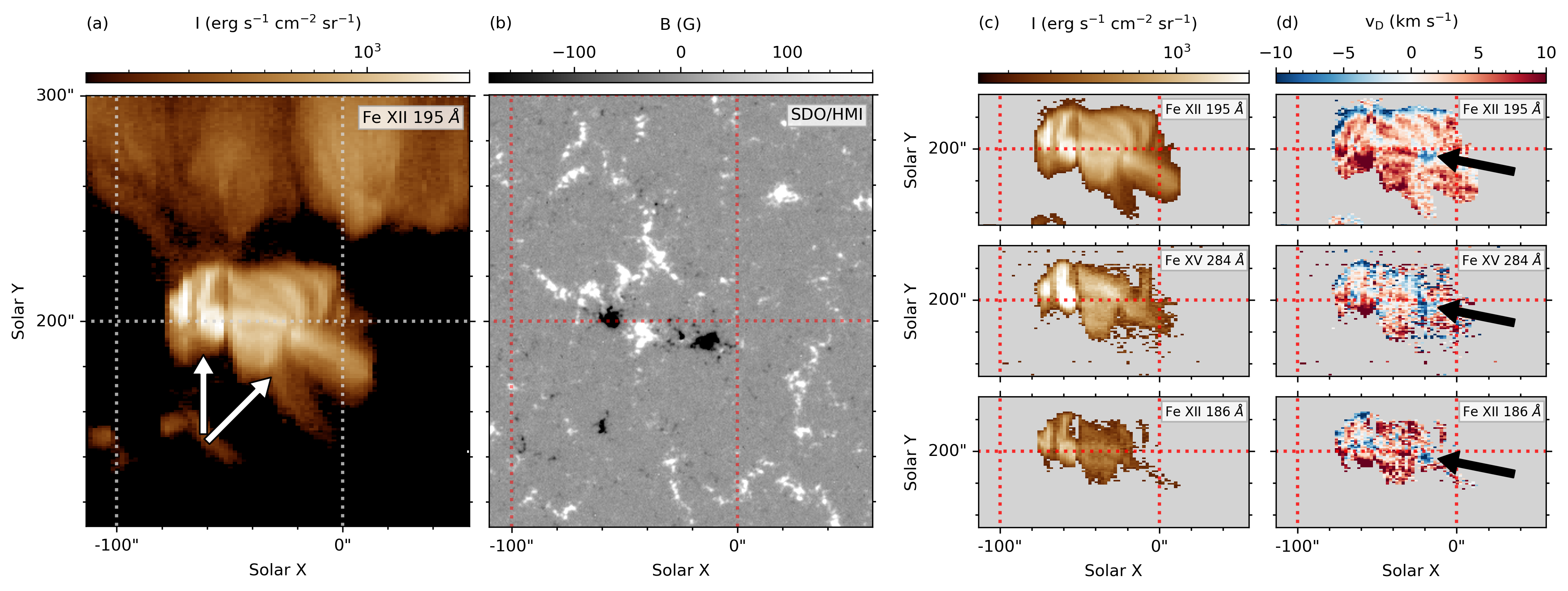}
   \caption{Observations of an equatorial coronal hole obtained on 25 June 2013.
   Panel (a): Hinode/EIS \FeXII\ intensity map with white arrows indicating the location of the two adjacent CBPs.
   Panel (b): SDO/HMI magnetogram.
   Panel (c): Zoomed-in intensity maps for the \FeXII, \FeXV, and \FeXIIb\ lines (from top to bottom), with values below 200 erg s$^{-1}$ cm$^{-1}$ sr$^{-1}$ masked in gray for clarity.
   Panel (d): Corresponding Doppler maps derived from Gaussian line fitting.
   The black arrows indicate the compact blueshifted region near the central parasitic polarity of the larger CBP.}
   \label{f:figure_05}
\end{figure*}

\subsection{Observational data}\label{subs:observations_data}
For the observational study, we used EIS \citep[][]{Culhane_etal:2007} on board the 
Hinode satellite \citep{Kosugi_etal:2007}. 
We searched the EIS archive for observations satisfying the following criteria:
\begin{itemize}
\item Proximity to disk center so that vertical flows are approximately aligned with the line of sight (LOS).
\item Observations obtained between 2011 and 2023, ensuring the availability of simultaneous SDO data.
\item Studies labeled as Harry P. Warren (HPW) or full special atlas. 
These studies contain at least the \FeXV, \FeXII, and \FeXIIb\ spectral lines, whose peak formation temperatures 
are approximately 2.2, 1.6, and 1.4 MK, respectively, comparable to the temperatures reached in the CBP simulation.
\item Requirement for raster-mode observations, since sit-and-stare observations are much less likely to have the slit intersect the relevant part of a CBP.
\end{itemize}
These criteria resulted in 573 EIS raster observations.
However, this sample cannot be used directly to estimate the occurrence rate of the upflows identified in our simulation.
For comparison, the CBP in our simulation has a diameter of approximately $14$~Mm, within the typical range of observed CBPs \citep{Madjarska:2019}, whereas the associated upflow patches reach maximum areas of only about $2$--$3$~Mm$^2$.
Even in the most favorable case of a roughly square feature, this corresponds to a linear size of only $1.4$--$1.7$ Mm, and would therefore remain unresolved by EIS, whose effective spatial resolution is about $2$--$3$~Mm on the Sun.
We consequently prioritized observations containing relatively large CBPs, thus mitigating the limitations imposed by the spatial resolution of 
EIS and maximizing the likelihood of detecting similar compact hot upflows to those identified in the simulation.
The observation described below provided the clearest candidate found in this search and was obtained between 23:26:16 UT on 25 June 2013 and 00:33:54 UT on 26 June 2013, 
targeting an equatorial coronal hole.
The observing sequence employed the \texttt{HPW022\_VEL\_480x512v1} study, using the $1"$ slit with a $2"$ step, and
included the \FeXV, \FeXII, and \FeXIIb\ emission lines.
The raster covered approximately $480"\times512"$ with 240 slit positions and a $15$~s exposure at each position.
We used the level-1 HDF5 data distributed by the Naval Research Laboratory (NRL) and 
\texttt{EISPAC}\footnote{\url{https://github.com/USNavalResearchLaboratory/eispac}}~\citep{Weberg_etal:2023} to fit the spectra with Gaussian profiles and derive line intensities, Doppler velocities, and line widths. 
Level-1 data were prepared by NRL to include corrections for dark current, cosmic rays, hot pixels, and CCD bias.  
Since EIS does not provide an absolute wavelength reference, the wavelength calibration includes corrections for the EIS slit tilt and thermally induced orbital spectral drift \citep{Kamio_etal:2010}. 
Residual wavelength offsets are further corrected in EISPAC by subtracting, at each raster position, the median Doppler velocity measured along the slit.
To provide the photospheric magnetic context for the EIS observations, we used a line-of-sight (LOS) magnetogram from the 
Helioseismic and Magnetic Imager \citep[HMI;][]{Scherrer_etal:2012} on board SDO obtained on 25 June 2013 at 23:26:15 UT.

\subsection{Observational results and comparison with the experiment}\label{subs:observations_results}
Figure~\ref{f:figure_05}a provides the observational context through an EIS \FeXII\ intensity map.
Two large adjacent CBPs are visible within the equatorial coronal hole: a larger CBP centered near
$(-25", 190")$ and a more compact one, but with higher  \FeXII\ intensity, near $(-70", 200")$.
The corresponding photospheric LOS magnetic field from HMI is shown in panel (b). 
Both CBPs are associated with negative parasitic polarities embedded in a predominantly positive magnetic environment.
Column (c) shows zoomed-in intensity maps of the \FeXII, \FeXV, and \FeXIIb\ lines, from top to bottom. 
Pixels with intensities below 200 erg s$^{-1}$ cm$^{-1}$ sr$^{-1}$ are masked in gray to restrict the analysis to regions 
with sufficient signal.
The corresponding Doppler maps are shown in column (d).
A compact blueshifted region is detected in the larger CBP, close to its central parasitic polarity (black arrows in 
the panels).
This signature is clearest in \FeXII, but a spatially similar blueshift is also present in \FeXV\ and \FeXIIb, although with 
a poorer signal-to-noise ratio. 
The location and compact morphology of this blueshifted region are consistent with the hot upflows found above the parasitic 
polarity in the numerical simulation.
The smaller CBP also appears to show a distinct upflow structure near its central parasitic polarity.
In this case, the signature is more apparent in \FeXV\ and \FeXIIb\ than in \FeXII.
These signatures, particularly that detected in the larger CBP, are suggestive of the upflows detected in the simulation, but they should be interpreted with caution. 
The residual uncertainty associated with the EIS wavelength-drift correction is about $4.4$~km~s$^{-1}$ \citep{Kamio_etal:2010}.
Moreover, the features lie close to the spatial and spectral resolution limits of EIS, and the signal-to-noise ratio is relatively 
poor in \FeXV\ and \FeXIIb. 
Higher-resolution coronal spectroscopy will therefore be required to establish their nature more conclusively (see discussion in Sect. \ref{subsub:discussion_upflows}).

\section{Conclusions and discussion}\label{s:discussion}
In this work we analyzed the self-consistent 3D radiation-magnetohydrodynamic model of a CBP, first presented by 
\citet{Nobrega-Siverio_etal:2023}, and show that recurrent episodes of strong heating occur in apparently unipolar
regions dominated by concentrated vertical magnetic field.
A detailed analysis of a representative case reveals that these events are related to 3D reconnection 
without nulls localized at heights $z\in[1.0,2.5]$~Mm near the CBP's positive parasitic
polarity. 
Local Joule heating per unit mass reaches values $Q_J/\rho \geq 2\times10^{13}$~erg g$^{-1}$ s$^{-1}$ at $z=1.5$~Mm, 
raising the plasma temperature to approximately $2$~MK.
The loops rooted in this region (Fig.~\ref{f:figure_02}) reach apex heights of at most $\approx6$~Mm and therefore remain 
well below the CBP null point, located near
$z=8$~Mm \citep{Nobrega-Siverio_etal:2023}.
This is relevant for the CBP heating: the heating of these loops is not associated with reconnection at the null, 
but instead results from Joule dissipation in a localized region of one of the loop footpoints.

The temperature increase in the analyzed region enhances the vertical gas-pressure gradient, driving upflows with speeds
ranging from $\geq 10$ to $50$~km s$^{-1}$.
The blueshifted response detected in Hinode/EIS observations close to the parasitic polarity of a CBP could well be the result of a phenomenon as described here.
In contrast, the horizontal response is mainly controlled by the Lorentz force, with horizontal velocities around a few
tens of km s$^{-1}$.

In the following, we organize the discussion into two main parts.
The first presents a systematic comparison with previous models of 3D
reconnection and coronal-loop heating, examining different properties
of the reconnection and related energy release (Sect.~\ref{subs:discussion_models}).
The second places our findings in an observational context, discussing
heating in apparently unipolar regions and the possible broader relevance
of the associated upflows (Sect.~\ref{subs:discussion_observations}).

\subsection{Comparison with other 3D models}\label{subs:discussion_models}
\subsubsection{The field-line integral of the parallel electric field, \texorpdfstring{$\Xi$}{Xi}} 
The field-line integral of the parallel electric field, $\Xi$, is a key
diagnostic of 3D magnetic reconnection away from magnetic nulls \citep[e.g.,][]{Schindler_etal:1988,Hesse_Schindler:1988,Priest_Forbes:2000,Biskamp:2005}.
In the representative event analyzed here, the largest values of $|\Xi|\approx10^5$~statV  (Fig.~\ref{f:figure_02}) 
imply a local magnetic-flux transfer rate of approximately $3\times10^{15}$~Mx~s$^{-1}$.
For reference, as an order-of-magnitude estimate, if a flux-transfer rate of this
magnitude were sustained, it would transfer an amount of magnetic flux
comparable to the total flux of a typical CBP,
$10^{19}$--$10^{20}$~Mx \citep{Madjarska:2019}, in approximately
1--9~hr.

The largest values of $|\Xi|$ we obtain are comparable to the peak local 3D reconnection rate inferred
from the stressed, multi-threaded coronal-loop simulations of
\cite{Reid_etal:2020}, performed with the Lare3D code
\citep{Arber_etal:2001}, which, from their figures, can be inferred to be around $4.6\times10^5$~statV.
Our result is also within the range obtained from 3D reconnection in the twisted-coronal-loop
simulations of \cite{Reale_etal:2016}, performed with the PLUTO code
\citep{Mignone_etal:2007,Mignone_etal:2012}, after correcting for an
apparent inconsistency in their current-density units 
\footnote{The values $\Xi\sim10^{13}$--$10^{15}$~statV quoted by
\citet{Reale_etal:2016} appear to result from labeling  $J_{\rm cr}=75$ as A~cm$^{-2}$ rather than
statA~cm$^{-2}$. Correcting for this factor yields 
$\Xi\sim3\times10^3$--$3\times10^5$~statV.}.
In the literature, we can find both larger and smaller values for $|\Xi|$ than those
reported here.
For instance,  \cite{Quinn_etal:2020} reported values of
$|\Xi|\sim3\times10^6$--$2\times10^7$~statV from 3D kink-unstable loop
simulations employing Lare3D.
At the lower end, \cite{Cozzo_etal:2026b} found
$|\Xi|\approx5\times10^2$~statV in narrow current sheets associated with
individual nanojets in their 3D coronal-loop simulations performed with the 
PLUTO code.
In general, small local $\Xi$ rates do not necessarily imply a weak thermal response, since numerous reconnection sites 
can operate intermittently and the efficiency of the heating also depends on the local density.

\subsubsection{The properties of the region: \texorpdfstring{$Q_\perp$}{Qperp}, \texorpdfstring{$\mathcal{T}_w$}{Tw}, \texorpdfstring{$\beta$}{beta}, and \texorpdfstring{$\theta_{JB}$}{thetajb} }
The fan-spine topology of the CBP naturally contains strong gradients
in field-line connectivity, leading to large values of the squashing factor $Q_\perp$.
In particular, field lines rooted close together around the
parasitic polarity can connect to widely separated regions along the fan.
Granular stressing further deforms this mapping, producing the complex
$Q_\perp$ network shown in Fig.~\ref{f:figure_02}g.
This complex QSL environment seems reminiscent of those obtained within the flux tubes in braiding simulations 
\citep[e.g.,][]{Pontin_Hornig:2015,Pontin_etal:2017,Pontin_etal:2026}.
However, we must recall that $Q_\perp$ is a topological quantity and does not by itself prove
that reconnection is taking place \citep[see the analysis by][]{Reid_etal:2020}.
As an example, Fig.~\ref{f:figure_02}g shows a region around $(x,y)\approx(18,20)$~Mm with $Q_\perp\gg1$; 
nevertheless, this region is not associated with either reconnection or heating.
Here, the signature of 3D reconnection is provided by the spatially nonuniform distribution of $|\Xi(x,y)|$, 
with local maxima, while $Q_J/\rho$ indicates whether the associated Joule heating is thermodynamically significant.

The modest values of the field-line twist number $\mathcal{T}_w\lesssim 1$ within the region of interest 
(Fig.~\ref{f:figure_02}f), together with the absence of a coherent high-twist flux-rope core, indicate that the reconnection 
event is unlikely to be triggered by the ideal kink instability.
The classical threshold for that instability corresponds approximately to $1.25$--$1.75$ turns for specific line-tied cylindrical and toroidal
flux ropes \citep{Hood_Priest:1981,Torok_etal:2004}.
Thus, the reconnection and heating event found around the parasitic polarity of the CBP differs from situations 
where kink instability in twisted loops produces an avalanche of reconnection and heating
\citep[e.g.,][]{Hood_etal:2009,Tam_etal:2015,Hood_etal:2016,
Reid_etal:2018,Reid_etal:2020,Cozzo_etal:2023}.

Finally, the reconnecting region is strongly dominated by the magnetic field, $\beta\ll1$ (see Fig.~\ref{f:figure_01}g),
and it is close to force-free, with $\theta_{JB}\approx1^\circ$
(see Fig.~\ref{f:figure_03}b).
This shows that thermodynamically significant 3D reconnection can occur in a locally nearly force-free magnetic environment. 
This is consistent with braided-field relaxation models, in which approximately force-free equilibria can contain 
thin current layers where reconnection takes place \citep{Pontin_Hornig:2015}.

\subsubsection{Low-atmosphere localization of the heating}
In our previous study \citep[][]{Nobrega-Siverio_etal:2023}, time averages over the 4.5~hr simulation showed that the 
strongest Joule and viscous heating in the CBP were concentrated at heights up to a few megameters above the photosphere, 
around the region dominated by the parasitic polarity.
Here, we identify the physical process underlying this time-averaged pattern: recurrent episodes of 3D reconnection 
occurring at $z\in[1.0,2.5]$~Mm near the loop footpoint rooted in the strong parasitic polarity (see 
Fig.~\ref{f:figure_02}a).
In addition, the magnetic geometry of the CBP loops anchored around the parasitic polarity 
favors energy dissipation at these low heights.
Near the parasitic polarity, the magnetic-field strength reaches up to $\approx250$~G at $z=1$~Mm, $\approx70$~G at $z=2.5$~Mm,
and $\approx30$~G at $z=4$~Mm.
Since the magnetic configuration in the region of interest is close to force-free and the current layers are nearly 
field-aligned (see Fig.~\ref{f:figure_03}b), $J$ approximately follows $B$, introducing a bias in the associated Joule heating 
($Q_J \propto J^2$) toward lower heights.
Such a concentration of energy release in the low corona is a recurrent result of models in which the magnetic field 
is stressed by photospheric motions \citep{Gudiksen_Nordlund:2005b,Hansteen_etal:2015}.
In contrast, straightened-loop models that neglect the large-scale expansion of the magnetic flux tube, such as the 
simulations of \cite{Breu_etal:2022} and \cite{Kuniyoshi_Morton:2026}, lack this geometrical tendency for $J^2$, 
and hence $Q_J$, to decrease from the footpoints toward the loop apex, which could affect the results about the heating
location.

\subsubsection{Reynolds and magnetic Prandtl numbers}
The much larger magnetic Reynolds number $R_m$ of the solar plasma is expected to modify the fine structure of our event.
The recent flux-braiding experiments by \cite{Pontin_etal:2026} show that increasing $R_m$ produces thinner, stronger, 
more numerous, and eventually fragmented current sheets, while the global reconnection rate remains 
only weakly dependent on $R_m$.
The reconnection region identified in the present paper may therefore represent a coarse-grained counterpart of a more intermittent and 
finely structured ensemble of real solar events.

The effective magnetic Prandtl numbers $P_m$ of the simulation also affect the partition of 
the released magnetic energy.
In the MURaM simulations of \cite{Rempel:2017} and \cite{Chen_etal:2022}, with an effective magnetic Prandtl number of
a few tens, magnetic energy is transferred mainly to plasma motions through Lorentz-force work and subsequently thermalized 
by viscous dissipation, whereas the event analyzed here is dominated by Joule heating.
This difference illustrates that the relative contributions of Joule and viscous heating are sensitive to the 
numerical treatment of the dissipation.
Nevertheless, the pressure-driven, field-aligned upflows reported here remain a physically plausible response to 
strong localized heating, regardless of whether the released energy is thermalized predominantly through
Joule or viscous dissipation.

\subsection{Observational perspective}\label{subs:discussion_observations}

\subsubsection{Heating in unipolar regions}
Even though our analysis is restricted to the parasitic polarity of a CBP, our findings on strong heating associated 
with 3D reconnection may be generic to situations involving predominantly unipolar magnetic fields, which is consistent 
with recent observational findings.
For instance, \cite{Tiwari_etal:2021} studied the polarities at the footpoints of 23 of the brightest coronal loops in 
an active region, finding that in roughly 40\% of the loop footpoints there is an absence of mixed-polarity 
magnetic flux observed with HMI. 
Similarly, utilizing the Microlensed Hyperspectral Imager \citep[MiHI;][]{van-Noort_etal:2022a,van-Noort_etal:2022b},
\cite{Chitta_etal:2024} found features resembling quiet-Sun Ellerman bombs in \Halpha\ (the authors refer to 
them as photospheric hot spots) in unipolar regions within HMI resolution.
More recently, \cite{Duan_etal:2025}, employing observations from the High Resolution 
Imager $\mathrm{HRI}_{\mathrm{EUV}}$~174 \AA\ \citep{Rochus_etal:2020} on board Solar Orbiter \citep{Muller_etal:2020},
reported that 24\% of small-scale EUV events
also occurred within unipolar magnetic fields as seen by HMI.
This suggests that reconnection between opposite polarities may not be not required to drive a substantial fraction of 
all these phenomena.

A legitimate concern is that the spatial resolution of HMI is limited compared with that of more recent instruments, 
such that apparently unipolar regions may in fact contain unresolved small-scale opposite-polarity magnetic flux.
Indeed, HMI has been shown to be unable to resolve small opposite-polarity magnetic elements in close proximity 
to unipolar regions \citep[e.g.,][]{Chitta_etal:2017,Raoufi_etal:2023}, to miss small-scale magnetic flux emergence 
\citep{Nobrega-Siverio_etal:2024}, and even to underestimate the magnetic flux in nominally 
unipolar regions \citep{Sinjan_etal:2024}. 
Nevertheless, even when high-resolution spectropolarimetric observations are considered, there is compelling evidence 
that magnetic reconnection operating within predominantly unipolar regions could play a significant role.
In particular, \cite{Judge_etal:2024a,Judge_Kuin:2024,Judge_etal:2024b} analyzed spectropolarimetric photospheric and 
chromospheric observations using the Visible Spectro-Polarimeter \citep[ViSP;][]{de-Wijn_etal:2022} at
the \textit{Daniel K. Inouye} Solar Telescope \citep[DKIST;][]{Rimmele_etal:2020} and the GRIS instrument \citep{Collados_etal:2012} 
at the GREGOR telescope \citep{Schmidt_etal:2012,Kleint_etal:2020}.
Their focus is on the coronal heating of long-lived, predominantly unipolar regions, with very little flux of 
opposite polarity.
They conclude that the observed bright chromospheric, transition-region, and coronal loops must be heated by 
mechanisms operating in unipolar magnetic fields, rather than by reconnection between opposite polarities.
In a similar fashion, through high-resolution magnetograms from the Swedish 1-m Solar Telescope 
\citep[SST;][]{Scharmer_etal:2003,Scharmer_etal:2019} using the CRisp Imaging SpectroPolarimeter instrument 
\citep[CRISP;][]{Scharmer_etal:2008}, many quiet-Sun Ellerman bombs have been reported within unipolar patches 
\citep[][]{JoshiJayant_etal:2020,Sand_etal:2025}.
Thus, our results support a picture in which 3D magnetic reconnection and heating can naturally occur in 
predominantly unipolar regions without requiring the presence of mixed-polarity configurations.

\subsubsection{Pressure-driven upflows following reconnection}\label{subsub:discussion_upflows}
The sequence found in our particular CBP experiment (localized 3D reconnection, efficient
heating, development of a pressure imbalance, and subsequent acceleration of plasma along 
the magnetic field) may be
general at magnetically stressed field lines, including those rooted in
predominantly unipolar regions.
In fact, recently, \cite{Zhu_etal:2025} investigated active region outflows using coordinated observations from 
Hinode and Solar Orbiter. 
They identified upflow regions located in areas that appear predominantly unipolar at the resolution of the 
Polarimetric and Helioseismic Imager \citep[PHI;][]{Solanki_etal:2020}.
Recent 3D numerical results by \cite{Bora_etal:2025} obtained with MURaM \citep[][and references therein]{Rempel:2017} 
also point in a similar direction.
These authors found significant hot upflows in a coronal-hole plume that originate near strongly structured QSLs, 
although the physical mechanism that initiates and accelerates the flows remains unresolved in that paper.
The sequence found in our model suggests a possible pathway in which localized reconnection heating at such interfaces 
generates a pressure imbalance that subsequently accelerates the plasma along the magnetic field.

The larger observed CBP shown in Fig.~\ref{f:figure_05} provides a potentially interesting
example of such a sequence.
Its compact blueshifted region is qualitatively consistent with the
simulated upflows, with both occurring close to the footpoint
rooted in the parasitic polarity.
This correspondence should, however, be regarded as suggestive
rather than conclusive.
For instance, the available observation consists of a single raster and
therefore provides no information about the temporal evolution of
the feature.
The measured Doppler shifts are comparable in magnitude to the residual velocity uncertainty 
of EIS \citep{Kamio_etal:2010}.
Moreover, the instrumental line width makes it difficult to
separate multiple unresolved velocity components, while the
signal-to-noise ratio in Fe~\textsc{xv} 284~\AA\ and
Fe~\textsc{xii} 186~\AA\ is insufficient for a robust
temperature-dependent analysis.
A larger statistical sample, together with time-dependent
spectroscopic observations, will be required to determine whether
such compact upflows are systematically associated with
predominantly unipolar CBP footpoints.
As mentioned in Sect.~\ref{s:observations}, constructing such a sample 
with EIS is particularly challenging because structures at the spatial 
scales found in our simulation are generally unresolved, while conventional 
slit spectroscopy also limits the simultaneous spatial and temporal coverage.
In this regard, the multi-slit approach of the Multi-slit Solar
Explorer \citep[MUSE;][]{De-Pontieu_etal:2020,
De-Pontieu_etal:2022,Cheung_etal:2022} will be particularly well suited to 
combining high spatial and temporal resolution over sufficiently large areas to survey and follow CBPs.
Together with Solar-C/EUVST \citep{Shimizu_etal:2020}, such observations should make it possible to establish statistically whether these compact upflows are a common response to localized heating at CBP footpoints.

\begin{acknowledgements}
This research has been supported by the European Research Council through the
Synergy Grant number 810218 (``The Whole Sun'', ERC-2018-SyG);
by the Centre National d'Etudes Spatiales (CNES) Solar Orbiter project;
by the Research Council of Norway (RCN) through its Centres of Excellence scheme, project number 262622;
and by the ESA Archival Research Visitor Programme.
This work is also part of grant CEX2025-001609-S, awarded to the Instituto de Astrofísica de Canarias under the Severo Ochoa Centre of Excellence program and funded by MICIU/AEI/10.13039/501100011033.
The authors acknowledge the computer resources at the MareNostrum
supercomputing installation and the technical support provided by the
Barcelona Supercomputing Center (BSC, RES-AECT-2021-1-0023,
RES-AECT-2022-2-0002), as well as the resources provided by 
Sigma2 - the National Infrastructure for High Performance Computing and Data Storage in Norway.
The use of UCAR's VAPOR software \citep{VAPOR_1,VAPOR_2} is gratefully acknowledged. 
The authors also thank Prof. Eric Priest for insightful discussions on 3D magnetic reconnection.
D.N.S. acknowledges the use of ChatGPT v5.6 to assist in optimizing the Python routines for data visualization as well
as refining sentence structures.
\end{acknowledgements}

\bibliographystyle{aa}
\bibliography{references}

\begin{thebibliography}{148}
\expandafter\ifx\csname natexlab\endcsname\relax\def\natexlab#1{#1}\fi

\bibitem[{{Alipour} \& {Safari}(2015)}]{Alipour_Safari:2015}
{Alipour}, N. \& {Safari}, H. 2015, \apj, 807, 175

\bibitem[{{Antolin} {et~al.}(2021){Antolin}, {Pagano}, {Testa}, {Petralia}, \&
  {Reale}}]{Antolin_etal:2021}
{Antolin}, P., {Pagano}, P., {Testa}, P., {Petralia}, A., \& {Reale}, F. 2021,
  Nature Astronomy, 5, 54

\bibitem[{{Arber} {et~al.}(2001){Arber}, {Longbottom}, {Gerrard}, \&
  {Milne}}]{Arber_etal:2001}
{Arber}, T.~D., {Longbottom}, A.~W., {Gerrard}, C.~L., \& {Milne}, A.~M. 2001,
  Journal of Computational Physics, 171, 151

\bibitem[{{Archontis} \& {Hood}(2013)}]{Archontis_Hood:2013}
{Archontis}, V. \& {Hood}, A.~W. 2013, \apjl, 769, L21

\bibitem[{{Aulanier} {et~al.}(2012){Aulanier}, {Janvier}, \&
  {Schmieder}}]{Aulanier_etal:2012}
{Aulanier}, G., {Janvier}, M., \& {Schmieder}, B. 2012, \aap, 543, A110

\bibitem[{Berger \& Prior(2006)}]{Berger_Prior_2006}
Berger, M.~A. \& Prior, C. 2006, Journal of Physics A: Mathematical and
  General, 39, 8321

\bibitem[{{Bhatnagar} {et~al.}(2025){Bhatnagar}, {Prasad}, {Rouppe van der
  Voort}, {N{\'o}brega-Siverio}, \& {Joshi}}]{Bhatnagar_etal:2025b}
{Bhatnagar}, A., {Prasad}, A., {Rouppe van der Voort}, L.,
  {N{\'o}brega-Siverio}, D., \& {Joshi}, J. 2025, \aap, 693, A221

\bibitem[{{Biskamp}(2005)}]{Biskamp:2005}
{Biskamp}, D. 2005, {Magnetic Reconnection in Plasmas} ({Cambridge University
  Press})

\bibitem[{{Bora} {et~al.}(2025){Bora}, {Chitta}, {Chen}, {Przybylski},
  {Pontin}, \& {Panyam}}]{Bora_etal:2025}
{Bora}, K., {Chitta}, L.~P., {Chen}, Y., {et~al.} 2025, \apjl, 994, L56

\bibitem[{{Bose} {et~al.}(2024){Bose}, {De Pontieu}, {Hansteen}, {Sainz Dalda},
  {Savage}, \& {Winebarger}}]{Bose_etal:2024}
{Bose}, S., {De Pontieu}, B., {Hansteen}, V., {et~al.} 2024, Nature Astronomy,
  8, 697

\bibitem[{{Breu} {et~al.}(2022){Breu}, {Peter}, {Cameron}, {Solanki},
  {Przybylski}, {Rempel}, \& {Chitta}}]{Breu_etal:2022}
{Breu}, C., {Peter}, H., {Cameron}, R., {et~al.} 2022, \aap, 658, A45

\bibitem[{{Carlsson} \& {Leenaarts}(2012)}]{Carlsson_Leenaarts:2012}
{Carlsson}, M. \& {Leenaarts}, J. 2012, \aap, 539, A39

\bibitem[{{Chen} {et~al.}(2022){Chen}, {Rempel}, \& {Fan}}]{Chen_etal:2022}
{Chen}, F., {Rempel}, M., \& {Fan}, Y. 2022, \apj, 937, 91

\bibitem[{{Chen} {et~al.}(2026){Chen}, {Peter}, {Przybylski}, {Chitta}, \&
  {Mandal}}]{Chen_etal:2026}
{Chen}, Y., {Peter}, H., {Przybylski}, D., {Chitta}, L.~P., \& {Mandal}, S.
  2026, \aap, 706, A141

\bibitem[{{Cheng} {et~al.}(2023){Cheng}, {Priest}, {Li}, {Chen}, {Aulanier},
  {Chitta}, {Wang}, {Peter}, {Zhu}, {Xing}, {Ding}, {Solanki}, {Berghmans},
  {Teriaca}, {Aznar Cuadrado}, {Zhukov}, {Guo}, {Long}, {Harra}, {Smith},
  {Rodriguez}, {Verbeeck}, {Barczynski}, \& {Parenti}}]{Cheng_etal:2023}
{Cheng}, X., {Priest}, E.~R., {Li}, H.~T., {et~al.} 2023, Nature
  Communications, 14, 2107

\bibitem[{{Cheung} {et~al.}(2022){Cheung}, {Mart{\'\i}nez-Sykora}, {Testa}, {De
  Pontieu}, {Chintzoglou}, {Rempel}, {Polito}, {Kerr}, {Reeves}, {Fletcher},
  {Jin}, {N{\'o}brega-Siverio}, {Danilovic}, {Antolin}, {Allred}, {Hansteen},
  {Ugarte-Urra}, {DeLuca}, {Longcope}, {Takasao}, {DeRosa}, {Boerner},
  {Jaeggli}, {Nitta}, {Daw}, {Carlsson}, {Golub}, \& {The}}]{Cheung_etal:2022}
{Cheung}, M. C.~M., {Mart{\'\i}nez-Sykora}, J., {Testa}, P., {et~al.} 2022,
  \apj, 926, 53

\bibitem[{{Chitta} {et~al.}(2022){Chitta}, {Peter}, {Parenti}, {Berghmans},
  {Auch{\`e}re}, {Solanki}, {Aznar Cuadrado}, {Sch{\"u}hle}, {Teriaca},
  {Mandal}, {Barczynski}, {Buchlin}, {Harra}, {Kraaikamp}, {Long}, {Rodriguez},
  {Schwanitz}, {Smith}, {Verbeeck}, {Zhukov}, {Liu}, \&
  {Cheung}}]{Chitta_etal:2022}
{Chitta}, L.~P., {Peter}, H., {Parenti}, S., {et~al.} 2022, \aap, 667, A166

\bibitem[{{Chitta} {et~al.}(2017){Chitta}, {Peter}, {Solanki}, {Barthol},
  {Gandorfer}, {Gizon}, {Hirzberger}, {Riethm{\"u}ller}, {van Noort}, {Blanco
  Rodr{\'\i}guez}, {Del Toro Iniesta}, {Orozco Su{\'a}rez}, {Schmidt},
  {Mart{\'\i}nez Pillet}, \& {Kn{\"o}lker}}]{Chitta_etal:2017}
{Chitta}, L.~P., {Peter}, H., {Solanki}, S.~K., {et~al.} 2017, \apjs, 229, 4

\bibitem[{{Chitta} {et~al.}(2024){Chitta}, {van Noort}, {Smitha}, {Priest}, \&
  {Rouppe van der Voort}}]{Chitta_etal:2024}
{Chitta}, L.~P., {van Noort}, M., {Smitha}, H.~N., {Priest}, E.~R., \& {Rouppe
  van der Voort}, L.~H.~M. 2024, \apj, 976, 134

\bibitem[{{Cirtain} {et~al.}(2013){Cirtain}, {Golub}, {Winebarger}, {De
  Pontieu}, {Kobayashi}, {Moore}, {Walsh}, {Korreck}, {Weber}, {McCauley},
  {Title}, {Kuzin}, \& {Deforest}}]{Cirtain_etal:2013}
{Cirtain}, J.~W., {Golub}, L., {Winebarger}, A.~R., {et~al.} 2013, \nat, 493,
  501

\bibitem[{{Collados} {et~al.}(2012){Collados}, {L{\'o}pez}, {P{\'a}ez},
  {Hern{\'a}ndez}, {Reyes}, {Calcines}, {Ballesteros}, {D{\'\i}az}, {Denker},
  {Lagg}, {Schlichenmaier}, {Schmidt}, {Solanki}, {Strassmeier}, {von der
  L{\"u}he}, \& {Volkmer}}]{Collados_etal:2012}
{Collados}, M., {L{\'o}pez}, R., {P{\'a}ez}, E., {et~al.} 2012, Astronomische
  Nachrichten, 333, 872

\bibitem[{{Cozzo} {et~al.}(2025){Cozzo}, {Pagano}, {Reale}, {Testa},
  {Petralia}, {Martinez-Sykora}, {Hansteen}, \& {De Pontieu}}]{Cozzo_etal:2025}
{Cozzo}, G., {Pagano}, P., {Reale}, F., {et~al.} 2025, \aap, 695, A40

\bibitem[{{Cozzo} {et~al.}(2023){Cozzo}, {Reid}, {Pagano}, {Reale}, \&
  {Hood}}]{Cozzo_etal:2023}
{Cozzo}, G., {Reid}, J., {Pagano}, P., {Reale}, F., \& {Hood}, A.~W. 2023,
  \aap, 678, A40

\bibitem[{{Cozzo} {et~al.}(2026{\natexlab{a}}){Cozzo}, {Testa},
  {Martinez-Sykora}, {Pagano}, {Reale}, {Rappazzo}, {Hansteen}, \& {De
  Pontieu}}]{Cozzo_etal:2026b}
{Cozzo}, G., {Testa}, P., {Martinez-Sykora}, J., {et~al.} 2026{\natexlab{a}},
  \apj, 998, 76

\bibitem[{{Cozzo} {et~al.}(2026{\natexlab{b}}){Cozzo}, {Testa},
  {Martinez-Sykora}, {Reale}, {Pagano}, {Rappazzo}, {Hansteen}, {De Pontieu},
  {Petralia}, {Alaimo}, {Fiorentino}, {D'Anca}, {Sciortino}, {Todaro}, {Lo
  Cicero}, \& {Barbera}}]{Cozzo_etal:2026a}
{Cozzo}, G., {Testa}, P., {Martinez-Sykora}, J., {et~al.} 2026{\natexlab{b}},
  \apj, 998, 75

\bibitem[{{Culhane} {et~al.}(2007){Culhane}, {Harra}, {James}, {Al-Janabi},
  {Bradley}, {Chaudry}, {Rees}, {Tandy}, {Thomas}, {Whillock}, {Winter},
  {Doschek}, {Korendyke}, {Brown}, {Myers}, {Mariska}, {Seely}, {Lang}, {Kent},
  {Shaughnessy}, {Young}, {Simnett}, {Castelli}, {Mahmoud}, {Mapson-Menard},
  {Probyn}, {Thomas}, {Davila}, {Dere}, {Windt}, {Shea}, {Hagood}, {Moye},
  {Hara}, {Watanabe}, {Matsuzaki}, {Kosugi}, {Hansteen}, \&
  {Wikstol}}]{Culhane_etal:2007}
{Culhane}, J.~L., {Harra}, L.~K., {James}, A.~M., {et~al.} 2007, \solphys, 243,
  19

\bibitem[{{Danilovic}(2017)}]{Danilovic_etal:2017}
{Danilovic}, S. 2017, \aap, 601, A122

\bibitem[{{De Pontieu} {et~al.}(2020){De Pontieu}, {Mart{\'\i}nez-Sykora},
  {Testa}, {Winebarger}, {Daw}, {Hansteen}, {Cheung}, \&
  {Antolin}}]{De-Pontieu_etal:2020}
{De Pontieu}, B., {Mart{\'\i}nez-Sykora}, J., {Testa}, P., {et~al.} 2020, \apj,
  888, 3

\bibitem[{{De Pontieu} {et~al.}(2022){De Pontieu}, {Testa},
  {Mart{\'\i}nez-Sykora}, {Antolin}, {Karampelas}, {Hansteen}, {Rempel},
  {Cheung}, {Reale}, {Danilovic}, {Pagano}, {Polito}, {De Moortel},
  {N{\'o}brega-Siverio}, {Van Doorsselaere}, {Petralia}, {Asgari-Targhi},
  {Boerner}, {Carlsson}, {Chintzoglou}, {Daw}, {DeLuca}, {Golub}, {Matsumoto},
  {Ugarte-Urra}, {McIntosh}, \& {the MUSE Team}}]{De-Pontieu_etal:2022}
{De Pontieu}, B., {Testa}, P., {Mart{\'\i}nez-Sykora}, J., {et~al.} 2022, \apj,
  926, 52

\bibitem[{{de Wijn} {et~al.}(2022){de Wijn}, {Casini}, {Carlile}, {Lecinski},
  {Sewell}, {Zmarzly}, {Eigenbrot}, {Beck}, {W{\"o}ger}, \&
  {Kn{\"o}lker}}]{de-Wijn_etal:2022}
{de Wijn}, A.~G., {Casini}, R., {Carlile}, A., {et~al.} 2022, \solphys, 297, 22

\bibitem[{{Duan} {et~al.}(2025){Duan}, {Chen}, {Hou}, {Sun}, \&
  {Shen}}]{Duan_etal:2025}
{Duan}, Y., {Chen}, H., {Hou}, Z., {Sun}, Z., \& {Shen}, Y. 2025, \apj, 979,
  195

\bibitem[{{Fletcher} {et~al.}(2011){Fletcher}, {Dennis}, {Hudson}, {Krucker},
  {Phillips}, {Veronig}, {Battaglia}, {Bone}, {Caspi}, {Chen}, {Gallagher},
  {Grigis}, {Ji}, {Liu}, {Milligan}, \& {Temmer}}]{Fletcher_etal:2011}
{Fletcher}, L., {Dennis}, B.~R., {Hudson}, H.~S., {et~al.} 2011, \ssr, 159, 19

\bibitem[{{Galsgaard} {et~al.}(2017){Galsgaard}, {Madjarska},
  {Moreno-Insertis}, {Huang}, \& {Wiegelmann}}]{Galsgaard_etal:2017}
{Galsgaard}, K., {Madjarska}, M.~S., {Moreno-Insertis}, F., {Huang}, Z., \&
  {Wiegelmann}, T. 2017, \aap, 606, A46

\bibitem[{{Golub} {et~al.}(1977){Golub}, {Krieger}, {Harvey}, \&
  {Vaiana}}]{Golub_etal:1977}
{Golub}, L., {Krieger}, A.~S., {Harvey}, J.~W., \& {Vaiana}, G.~S. 1977,
  \solphys, 53, 111

\bibitem[{{Go{\v{s}}i{\'c}} {et~al.}(2018){Go{\v{s}}i{\'c}}, {de la Cruz
  Rodr{\'\i}guez}, {De Pontieu}, {Bellot Rubio}, {Carlsson}, {Esteban Pozuelo},
  {Ortiz}, \& {Polito}}]{Gosic_etal:2018}
{Go{\v{s}}i{\'c}}, M., {de la Cruz Rodr{\'\i}guez}, J., {De Pontieu}, B.,
  {et~al.} 2018, \apj, 857, 48

\bibitem[{{Green} {et~al.}(2011){Green}, {Kliem}, \&
  {Wallace}}]{Green_Wallace:2011}
{Green}, L.~M., {Kliem}, B., \& {Wallace}, A.~J. 2011, \aap, 526, A2

\bibitem[{{Gudiksen} {et~al.}(2011){Gudiksen}, {Carlsson}, {Hansteen}, {Hayek},
  {Leenaarts}, \& {Mart{\'\i}nez-Sykora}}]{Gudiksen_etal:2011}
{Gudiksen}, B.~V., {Carlsson}, M., {Hansteen}, V.~H., {et~al.} 2011, \aap, 531,
  A154

\bibitem[{{Gudiksen} \&
  {Nordlund}(2005{\natexlab{a}})}]{Gudiksen_Nordlund:2005b}
{Gudiksen}, B.~V. \& {Nordlund}, {\r{A}}. 2005{\natexlab{a}}, \apj, 618, 1031

\bibitem[{{Gudiksen} \&
  {Nordlund}(2005{\natexlab{b}})}]{Gudiksen_Nordlund:2005a}
{Gudiksen}, B.~V. \& {Nordlund}, {\r{A}}. 2005{\natexlab{b}}, \apj, 618, 1020

\bibitem[{{Hansteen} {et~al.}(2015){Hansteen}, {Guerreiro}, {De Pontieu}, \&
  {Carlsson}}]{Hansteen_etal:2015}
{Hansteen}, V., {Guerreiro}, N., {De Pontieu}, B., \& {Carlsson}, M. 2015,
  \apj, 811, 106

\bibitem[{{Hansteen} {et~al.}(2019){Hansteen}, {Ortiz}, {Archontis},
  {Carlsson}, {Pereira}, \& {Bj{\o}rgen}}]{Hansteen_etal:2019}
{Hansteen}, V., {Ortiz}, A., {Archontis}, V., {et~al.} 2019, \aap, 626, A33

\bibitem[{{Hayek} {et~al.}(2010){Hayek}, {Asplund}, {Carlsson}, {Trampedach},
  {Collet}, {Gudiksen}, {Hansteen}, \& {Leenaarts}}]{Hayek_etal:2010}
{Hayek}, W., {Asplund}, M., {Carlsson}, M., {et~al.} 2010, \aap, 517, A49

\bibitem[{{Hesse} \& {Birn}(1993)}]{Hesse_Birn:1993}
{Hesse}, M. \& {Birn}, J. 1993, Advances in Space Research, 13, 249

\bibitem[{{Hesse} {et~al.}(2005){Hesse}, {Forbes}, \& {Birn}}]{Hesse_etal:2005}
{Hesse}, M., {Forbes}, T.~G., \& {Birn}, J. 2005, \apj, 631, 1227

\bibitem[{{Hesse} \& {Schindler}(1988)}]{Hesse_Schindler:1988}
{Hesse}, M. \& {Schindler}, K. 1988, \jgr, 93, 5559

\bibitem[{{Hood} {et~al.}(2009){Hood}, {Browning}, \& {van der
  Linden}}]{Hood_etal:2009}
{Hood}, A.~W., {Browning}, P.~K., \& {van der Linden}, R.~A.~M. 2009, \aap,
  506, 913

\bibitem[{{Hood} {et~al.}(2016){Hood}, {Cargill}, {Browning}, \&
  {Tam}}]{Hood_etal:2016}
{Hood}, A.~W., {Cargill}, P.~J., {Browning}, P.~K., \& {Tam}, K.~V. 2016, \apj,
  817, 5

\bibitem[{{Hood} \& {Priest}(1981)}]{Hood_Priest:1981}
{Hood}, A.~W. \& {Priest}, E.~R. 1981, Geophysical and Astrophysical Fluid
  Dynamics, 17, 297

\bibitem[{{Janvier} {et~al.}(2013){Janvier}, {Aulanier}, {Pariat}, \&
  {D{\'e}moulin}}]{Janvier_etal:2013}
{Janvier}, M., {Aulanier}, G., {Pariat}, E., \& {D{\'e}moulin}, P. 2013, \aap,
  555, A77

\bibitem[{{Joshi} {et~al.}(2020{\natexlab{a}}){Joshi}, {Rouppe van der Voort},
  \& {de la Cruz Rodr{\'\i}guez}}]{JoshiJayant_etal:2020}
{Joshi}, J., {Rouppe van der Voort}, L. H.~M., \& {de la Cruz Rodr{\'\i}guez},
  J. 2020{\natexlab{a}}, \aap, 641, L5

\bibitem[{{Joshi} {et~al.}(2020{\natexlab{b}}){Joshi}, {Chandra}, {Schmieder},
  {Moreno-Insertis}, {Aulanier}, {N{\'o}brega-Siverio}, \&
  {Devi}}]{JoshiR_etal:2020}
{Joshi}, R., {Chandra}, R., {Schmieder}, B., {et~al.} 2020{\natexlab{b}}, \aap,
  639, A22

\bibitem[{{Judge} \& {Ionson}(2024)}]{Judge_Ionson:2024}
{Judge}, P. \& {Ionson}, J.~A., eds. 2024, Astrophysics and Space Science
  Library, Vol. 470, {The Problem of Coronal Heating: A Rosetta Stone for
  Electrodynamic Coupling in Cosmic Plasmas}

\bibitem[{{Judge} {et~al.}(2024{\natexlab{a}}){Judge}, {Kleint}, {Casini}, {de
  Wijn}, {Schad}, \& {Tritschler}}]{Judge_etal:2024a}
{Judge}, P., {Kleint}, L., {Casini}, R., {et~al.} 2024{\natexlab{a}}, \apj,
  960, 129

\bibitem[{{Judge} {et~al.}(2024{\natexlab{b}}){Judge}, {Kleint}, \&
  {Kuckein}}]{Judge_etal:2024b}
{Judge}, P.~G., {Kleint}, L., \& {Kuckein}, C. 2024{\natexlab{b}}, \apj, 970,
  147

\bibitem[{{Judge} \& {Kuin}(2024)}]{Judge_Kuin:2024}
{Judge}, P.~G. \& {Kuin}, N. P.~M. 2024, \apj, 970, 130

\bibitem[{{Kamio} {et~al.}(2010){Kamio}, {Hara}, {Watanabe}, {Fredvik}, \&
  {Hansteen}}]{Kamio_etal:2010}
{Kamio}, S., {Hara}, H., {Watanabe}, T., {Fredvik}, T., \& {Hansteen}, V.~H.
  2010, \solphys, 266, 209

\bibitem[{{Kleint} {et~al.}(2020){Kleint}, {Berkefeld}, {Esteves}, {Sonner},
  {Volkmer}, {Gerber}, {Kr{\"a}mer}, {Grassin}, \&
  {Berdyugina}}]{Kleint_etal:2020}
{Kleint}, L., {Berkefeld}, T., {Esteves}, M., {et~al.} 2020, \aap, 641, A27

\bibitem[{{Klimchuk}(2006)}]{Klimchuk:2006}
{Klimchuk}, J.~A. 2006, \solphys, 234, 41

\bibitem[{{Kosugi} {et~al.}(2007){Kosugi}, {Matsuzaki}, {Sakao}, {Shimizu},
  {Sone}, {Tachikawa}, {Hashimoto}, {Minesugi}, {Ohnishi}, {Yamada}, {Tsuneta},
  {Hara}, {Ichimoto}, {Suematsu}, {Shimojo}, {Watanabe}, {Shimada}, {Davis},
  {Hill}, {Owens}, {Title}, {Culhane}, {Harra}, {Doschek}, \&
  {Golub}}]{Kosugi_etal:2007}
{Kosugi}, T., {Matsuzaki}, K., {Sakao}, T., {et~al.} 2007, \solphys, 243, 3

\bibitem[{{Kraus} {et~al.}(2023){Kraus}, {Bourdin}, {Zender}, {Bergmann}, \&
  {Hanslmeier}}]{Kraus_etal:2023}
{Kraus}, I., {Bourdin}, P.-A., {Zender}, J., {Bergmann}, M., \& {Hanslmeier},
  A. 2023, \aap, 678, A184

\bibitem[{{Kraus} {et~al.}(2025){Kraus}, {Bourdin}, {Zender}, {Bergmann}, \&
  {Hanslmeier}}]{Kraus_etal:2025}
{Kraus}, I., {Bourdin}, P.-A., {Zender}, J., {Bergmann}, M., \& {Hanslmeier},
  A. 2025, \aap, 694, A240

\bibitem[{{Kumar} {et~al.}(2019{\natexlab{a}}){Kumar}, {Karpen}, {Antiochos},
  {Wyper}, \& {DeVore}}]{Kumar_etal:2019b}
{Kumar}, P., {Karpen}, J.~T., {Antiochos}, S.~K., {Wyper}, P.~F., \& {DeVore},
  C.~R. 2019{\natexlab{a}}, \apjl, 885, L15

\bibitem[{{Kumar} {et~al.}(2018){Kumar}, {Karpen}, {Antiochos}, {Wyper},
  {DeVore}, \& {DeForest}}]{Kumar_etal:2018}
{Kumar}, P., {Karpen}, J.~T., {Antiochos}, S.~K., {et~al.} 2018, \apj, 854, 155

\bibitem[{{Kumar} {et~al.}(2019{\natexlab{b}}){Kumar}, {Karpen}, {Antiochos},
  {Wyper}, {DeVore}, \& {DeForest}}]{Kumar_etal:2019a}
{Kumar}, P., {Karpen}, J.~T., {Antiochos}, S.~K., {et~al.} 2019{\natexlab{b}},
  \apj, 873, 93

\bibitem[{{Kuniyoshi} \& {Morton}(2026)}]{Kuniyoshi_Morton:2026}
{Kuniyoshi}, H. \& {Morton}, R.~J. 2026, \apj, 1009, 129

\bibitem[{{Lemen} {et~al.}(2012){Lemen}, {Title}, {Akin}, {Boerner}, {Chou},
  {Drake}, {Duncan}, {Edwards}, {Friedlaender}, {Heyman}, {Hurlburt}, {Katz},
  {Kushner}, {Levay}, {Lindgren}, {Mathur}, {McFeaters}, {Mitchell}, {Rehse},
  {Schrijver}, {Springer}, {Stern}, {Tarbell}, {Wuelser}, {Wolfson}, {Yanari},
  {Bookbinder}, {Cheimets}, {Caldwell}, {Deluca}, {Gates}, {Golub}, {Park},
  {Podgorski}, {Bush}, {Scherrer}, {Gummin}, {Smith}, {Auker}, {Jerram},
  {Pool}, {Soufli}, {Windt}, {Beardsley}, {Clapp}, {Lang}, \&
  {Waltham}}]{Lemen_etal:2012}
{Lemen}, J.~R., {Title}, A.~M., {Akin}, D.~J., {et~al.} 2012, \solphys, 275, 17

\bibitem[{Li {et~al.}(2019)Li, Jaroszynski, Pearse, Orf, \& Clyne}]{VAPOR_1}
Li, S., Jaroszynski, S., Pearse, S., Orf, L., \& Clyne, J. 2019, Atmosphere,
  10, 488

\bibitem[{{Liu} {et~al.}(2018){Liu}, {Erd{\'e}lyi}, {Wang}, \&
  {Liu}}]{Liu_etal:2018}
{Liu}, J., {Erd{\'e}lyi}, R., {Wang}, Y., \& {Liu}, R. 2018, \apj, 852, 10

\bibitem[{{Liu} {et~al.}(2016){Liu}, {Kliem}, {Titov}, {Chen}, {Wang}, {Wang},
  {Liu}, {Xu}, \& {Wiegelmann}}]{Liu_etal:2016}
{Liu}, R., {Kliem}, B., {Titov}, V.~S., {et~al.} 2016, \apj, 818, 148

\bibitem[{{Madjarska}(2019)}]{Madjarska:2019}
{Madjarska}, M.~S. 2019, Living Reviews in Solar Physics, 16, 2

\bibitem[{{Madjarska} {et~al.}(2021){Madjarska}, {Chae}, {Moreno-Insertis},
  {Hou}, {N{\'o}brega-Siverio}, {Kwak}, {Galsgaard}, \&
  {Cho}}]{Madjarska_etal:2021}
{Madjarska}, M.~S., {Chae}, J., {Moreno-Insertis}, F., {et~al.} 2021, \aap,
  646, A107

\bibitem[{{Mignone} {et~al.}(2007){Mignone}, {Bodo}, {Massaglia}, {Matsakos},
  {Tesileanu}, {Zanni}, \& {Ferrari}}]{Mignone_etal:2007}
{Mignone}, A., {Bodo}, G., {Massaglia}, S., {et~al.} 2007, \apjs, 170, 228

\bibitem[{{Mignone} {et~al.}(2012){Mignone}, {Zanni}, {Tzeferacos}, {van
  Straalen}, {Colella}, \& {Bodo}}]{Mignone_etal:2012}
{Mignone}, A., {Zanni}, C., {Tzeferacos}, P., {et~al.} 2012, \apjs, 198, 7

\bibitem[{{Mondal} {et~al.}(2023){Mondal}, {Klimchuk}, {Vadawale}, {Sarkar},
  {Del Zanna}, {Athiray}, {Mithun}, {Mason}, \& {Bhardwaj}}]{Mondal_etal:2023}
{Mondal}, B., {Klimchuk}, J.~A., {Vadawale}, S.~V., {et~al.} 2023, \apj, 945,
  37

\bibitem[{{Moreno-Insertis} \&
  {Galsgaard}(2013)}]{Moreno-Insertis_Galsgaard:2013}
{Moreno-Insertis}, F. \& {Galsgaard}, K. 2013, \apj, 771, 20

\bibitem[{{Moreno-Insertis} {et~al.}(2008){Moreno-Insertis}, {Galsgaard}, \&
  {Ugarte-Urra}}]{Moreno-Insertis_etal:2008}
{Moreno-Insertis}, F., {Galsgaard}, K., \& {Ugarte-Urra}, I. 2008, \apjl, 673,
  L211

\bibitem[{{Moreno-Insertis} {et~al.}(2025){Moreno-Insertis}, {Hansteen}, \&
  {N{\'o}brega-Siverio}}]{Moreno-Insertis_etal:2025}
{Moreno-Insertis}, F., {Hansteen}, V.~H., \& {N{\'o}brega-Siverio}, D. 2025,
  \apj, 995, 137

\bibitem[{{Mou} {et~al.}(2016){Mou}, {Huang}, {Xia}, {Madjarska}, {Li}, {Fu},
  {Jiao}, \& {Hou}}]{Mou_etal:2016}
{Mou}, C., {Huang}, Z., {Xia}, L., {et~al.} 2016, \apj, 818, 9

\bibitem[{{M{\"u}ller} {et~al.}(2020){M{\"u}ller}, {St. Cyr}, {Zouganelis},
  {Gilbert}, {Marsden}, {Nieves-Chinchilla}, {Antonucci}, {Auch{\`e}re},
  {Berghmans}, {Horbury}, {Howard}, {Krucker}, {Maksimovic}, {Owen}, {Rochus},
  {Rodriguez-Pacheco}, {Romoli}, {Solanki}, {Bruno}, {Carlsson}, {Fludra},
  {Harra}, {Hassler}, {Livi}, {Louarn}, {Peter}, {Sch{\"u}hle}, {Teriaca}, {del
  Toro Iniesta}, {Wimmer-Schweingruber}, {Marsch}, {Velli}, {De Groof},
  {Walsh}, \& {Williams}}]{Muller_etal:2020}
{M{\"u}ller}, D., {St. Cyr}, O.~C., {Zouganelis}, I., {et~al.} 2020, \aap, 642,
  A1

\bibitem[{{Ni} {et~al.}(2022){Ni}, {Cheng}, \& {Lin}}]{Ni_etal:2022}
{Ni}, L., {Cheng}, G., \& {Lin}, J. 2022, \aap, 665, A116

\bibitem[{{N{\'o}brega-Siverio} {et~al.}(2024){N{\'o}brega-Siverio}, {Cabello},
  {Bose}, {Rouppe van der Voort}, {Joshi}, {Froment}, \&
  {Henriques}}]{Nobrega-Siverio_etal:2024}
{N{\'o}brega-Siverio}, D., {Cabello}, I., {Bose}, S., {et~al.} 2024, \aap, 686,
  A218

\bibitem[{{N{\'o}brega-Siverio} {et~al.}(2025){N{\'o}brega-Siverio}, {Joshi},
  {Sola-Viladesau}, {Berghmans}, \& {Lim}}]{Nobrega-Siverio_etal:2025}
{N{\'o}brega-Siverio}, D., {Joshi}, R., {Sola-Viladesau}, E., {Berghmans}, D.,
  \& {Lim}, D. 2025, \aap, 702, A188

\bibitem[{{N{\'o}brega-Siverio} {et~al.}(2023){N{\'o}brega-Siverio},
  {Moreno-Insertis}, {Galsgaard}, {Krikova}, {Rouppe van der Voort}, {Joshi},
  \& {Madjarska}}]{Nobrega-Siverio_etal:2023}
{N{\'o}brega-Siverio}, D., {Moreno-Insertis}, F., {Galsgaard}, K., {et~al.}
  2023, \apjl, 958, L38

\bibitem[{{N{\'o}brega-Siverio} {et~al.}(2016){N{\'o}brega-Siverio},
  {Moreno-Insertis}, \& {Mart{\'\i}nez-Sykora}}]{Nobrega-Siverio_etal:2016}
{N{\'o}brega-Siverio}, D., {Moreno-Insertis}, F., \& {Mart{\'\i}nez-Sykora}, J.
  2016, \apj, 822, 18

\bibitem[{{Norton} {et~al.}(2006){Norton}, {Graham}, {Ulrich}, {Schou},
  {Tomczyk}, {Liu}, {Lites}, {Ariste}, {Bush}, {Socas-Navarro}, \&
  {Scherrer}}]{Norton_etal:2006}
{Norton}, A.~A., {Graham}, J.~P., {Ulrich}, R.~K., {et~al.} 2006, \solphys,
  239, 69

\bibitem[{{Ortiz} {et~al.}(2020){Ortiz}, {Hansteen}, {N{\'o}brega-Siverio}, \&
  {Rouppe van der Voort}}]{Ortiz_etal:2020}
{Ortiz}, A., {Hansteen}, V.~H., {N{\'o}brega-Siverio}, D., \& {Rouppe van der
  Voort}, L. 2020, \aap, 633, A58

\bibitem[{{Panesar} {et~al.}(2017){Panesar}, {Sterling}, \&
  {Moore}}]{Panesar_etal:2017}
{Panesar}, N.~K., {Sterling}, A.~C., \& {Moore}, R.~L. 2017, \apj, 844, 131

\bibitem[{{Panesar} {et~al.}(2016){Panesar}, {Sterling}, {Moore}, \&
  {Chakrapani}}]{Panesar_etal:2016}
{Panesar}, N.~K., {Sterling}, A.~C., {Moore}, R.~L., \& {Chakrapani}, P. 2016,
  \apjl, 832, L7

\bibitem[{{Pariat} {et~al.}(2009){Pariat}, {Antiochos}, \&
  {DeVore}}]{Pariat_etal:2009}
{Pariat}, E., {Antiochos}, S.~K., \& {DeVore}, C.~R. 2009, \apj, 691, 61

\bibitem[{{Parker}(1972)}]{Parker:1972}
{Parker}, E.~N. 1972, \apj, 174, 499

\bibitem[{{Parnell} \& {De Moortel}(2012)}]{Parnell_De-Moortel:2012}
{Parnell}, C.~E. \& {De Moortel}, I. 2012, Philosophical Transactions of the
  Royal Society of London Series A, 370, 3217

\bibitem[{{Patsourakos} \& {Archontis}(2025)}]{Patsourakos_Archontis:2025}
{Patsourakos}, S. \& {Archontis}, V. 2025, \aap, 699, A87

\bibitem[{Pearse {et~al.}(2023)Pearse, Li, clyne, StasJ, CoreCode, Daves,
  Hallock, Eroglu, Poplawski, \& Lacroix}]{VAPOR_2}
Pearse, S., Li, S., clyne, {et~al.} 2023, Zenodo

\bibitem[{{Pesnell} {et~al.}(2012){Pesnell}, {Thompson}, \&
  {Chamberlin}}]{Pesnell_etal:2012}
{Pesnell}, W.~D., {Thompson}, B.~J., \& {Chamberlin}, P.~C. 2012, \solphys,
  275, 3

\bibitem[{{Peter} {et~al.}(2019){Peter}, {Huang}, {Chitta}, \&
  {Young}}]{Peter_etal:2019}
{Peter}, H., {Huang}, Y.-M., {Chitta}, L.~P., \& {Young}, P.~R. 2019, \aap,
  628, A8

\bibitem[{{Petrova} {et~al.}(2024){Petrova}, {Van Doorsselaere}, {Berghmans},
  {Parenti}, {Valori}, \& {Plowman}}]{Petrova_etal:2024}
{Petrova}, E., {Van Doorsselaere}, T., {Berghmans}, D., {et~al.} 2024, \aap,
  687, A13

\bibitem[{{Pontin} {et~al.}(2026){Pontin}, {Galsgaard}, \&
  {Klimchuk}}]{Pontin_etal:2026}
{Pontin}, D.~I., {Galsgaard}, K., \& {Klimchuk}, J.~A. 2026, \apj, 1006, 50

\bibitem[{{Pontin} \& {Hornig}(2015)}]{Pontin_Hornig:2015}
{Pontin}, D.~I. \& {Hornig}, G. 2015, \apj, 805, 47

\bibitem[{{Pontin} {et~al.}(2017){Pontin}, {Janvier}, {Tiwari}, {Galsgaard},
  {Winebarger}, \& {Cirtain}}]{Pontin_etal:2017}
{Pontin}, D.~I., {Janvier}, M., {Tiwari}, S.~K., {et~al.} 2017, \apj, 837, 108

\bibitem[{{Priest} \& {Forbes}(2000)}]{Priest_Forbes:2000}
{Priest}, E. \& {Forbes}, T. 2000, {Magnetic Reconnection: MHD Theory and
  Applications} ({Cambridge University Press})

\bibitem[{{Priest} {et~al.}(2018){Priest}, {Chitta}, \&
  {Syntelis}}]{Priest_etal:2018}
{Priest}, E.~R., {Chitta}, L.~P., \& {Syntelis}, P. 2018, \apjl, 862, L24

\bibitem[{{Priest} \& {D{\'e}moulin}(1995)}]{Priest_Demoulin:1995}
{Priest}, E.~R. \& {D{\'e}moulin}, P. 1995, \jgr, 100, 23443

\bibitem[{{Priest} {et~al.}(2002){Priest}, {Heyvaerts}, \&
  {Title}}]{Priest_etal:2002}
{Priest}, E.~R., {Heyvaerts}, J.~F., \& {Title}, A.~M. 2002, \apj, 576, 533

\bibitem[{{Priest} {et~al.}(1994){Priest}, {Parnell}, \&
  {Martin}}]{Priest_etal:1994}
{Priest}, E.~R., {Parnell}, C.~E., \& {Martin}, S.~F. 1994, \apj, 427, 459

\bibitem[{{Quinn} {et~al.}(2020){Quinn}, {MacTaggart}, \&
  {Simitev}}]{Quinn_etal:2020}
{Quinn}, J., {MacTaggart}, D., \& {Simitev}, R.~D. 2020, Communications in
  Nonlinear Science and Numerical Simulations, 83, 105131

\bibitem[{{Raouafi} {et~al.}(2023){Raouafi}, {Stenborg}, {Seaton}, {Wang},
  {Wang}, {DeForest}, {Bale}, {Drake}, {Uritsky}, {Karpen}, {DeVore},
  {Sterling}, {Horbury}, {Harra}, {Bourouaine}, {Kasper}, {Kumar}, {Phan}, \&
  {Velli}}]{Raoufi_etal:2023}
{Raouafi}, N.~E., {Stenborg}, G., {Seaton}, D.~B., {et~al.} 2023, \apj, 945, 28

\bibitem[{{Rappazzo} {et~al.}(2007){Rappazzo}, {Velli}, {Einaudi}, \&
  {Dahlburg}}]{Rappazzo_etal:2007}
{Rappazzo}, A.~F., {Velli}, M., {Einaudi}, G., \& {Dahlburg}, R.~B. 2007,
  \apjl, 657, L47

\bibitem[{{Reale} {et~al.}(2016){Reale}, {Orlando}, {Guarrasi}, {Mignone},
  {Peres}, {Hood}, \& {Priest}}]{Reale_etal:2016}
{Reale}, F., {Orlando}, S., {Guarrasi}, M., {et~al.} 2016, \apj, 830, 21

\bibitem[{{Reid} {et~al.}(2016){Reid}, {Mathioudakis}, {Doyle}, {Scullion},
  {Nelson}, {Henriques}, \& {Ray}}]{Reid_etal:2016}
{Reid}, A., {Mathioudakis}, M., {Doyle}, J.~G., {et~al.} 2016, \apj, 823, 110

\bibitem[{{Reid} {et~al.}(2018){Reid}, {Hood}, {Parnell}, {Browning}, \&
  {Cargill}}]{Reid_etal:2018}
{Reid}, J., {Hood}, A.~W., {Parnell}, C.~E., {Browning}, P.~K., \& {Cargill},
  P.~J. 2018, \aap, 615, A84

\bibitem[{{Reid} {et~al.}(2020){Reid}, {Parnell}, {Hood}, \&
  {Browning}}]{Reid_etal:2020}
{Reid}, J., {Parnell}, C.~E., {Hood}, A.~W., \& {Browning}, P.~K. 2020, \aap,
  633, A92

\bibitem[{{Rempel}(2017)}]{Rempel:2017}
{Rempel}, M. 2017, \apj, 834, 10

\bibitem[{{Rimmele} {et~al.}(2020){Rimmele}, {Warner}, {Keil}, {Goode},
  {Kn{\"o}lker}, {Kuhn}, {Rosner}, {McMullin}, {Casini}, {Lin}, {W{\"o}ger},
  {von der L{\"u}he}, {Tritschler}, {Davey}, {de Wijn}, {Elmore}, {Fehlmann},
  {Harrington}, {Jaeggli}, {Rast}, {Schad}, {Schmidt}, {Mathioudakis},
  {Mickey}, {Anan}, {Beck}, {Marshall}, {Jeffers}, {Oschmann}, {Beard},
  {Berst}, {Cowan}, {Craig}, {Cross}, {Cummings}, {Donnelly}, {de Vanssay},
  {Eigenbrot}, {Ferayorni}, {Foster}, {Galapon}, {Gedrites}, {Gonzales},
  {Goodrich}, {Gregory}, {Guzman}, {Guzzo}, {Hegwer}, {Hubbard}, {Hubbard},
  {Johansson}, {Johnson}, {Liang}, {Liang}, {McQuillen}, {Mayer}, {Newman},
  {Onodera}, {Phelps}, {Puentes}, {Richards}, {Rimmele}, {Sekulic}, {Shimko},
  {Simison}, {Smith}, {Starman}, {Sueoka}, {Summers}, {Szabo}, {Szabo},
  {Wampler}, {Williams}, \& {White}}]{Rimmele_etal:2020}
{Rimmele}, T.~R., {Warner}, M., {Keil}, S.~L., {et~al.} 2020, \solphys, 295,
  172

\bibitem[{{Rochus} {et~al.}(2020){Rochus}, {Auch{\`e}re}, {Berghmans}, {Harra},
  {Schmutz}, {Sch{\"u}hle}, {Addison}, {Appourchaux}, {Aznar Cuadrado},
  {Baker}, {Barbay}, {Bates}, {BenMoussa}, {Bergmann}, {Beurthe}, {Borgo},
  {Bonte}, {Bouzit}, {Bradley}, {B{\"u}chel}, {Buchlin}, {B{\"u}chner},
  {Cab{\'e}}, {Cadiergues}, {Chaigneau}, {Chares}, {Choque Cortez}, {Coker},
  {Condamin}, {Coumar}, {Curdt}, {Cutler}, {Davies}, {Davison}, {Defise}, {Del
  Zanna}, {Delmotte}, {Delouille}, {Dolla}, {Dumesnil}, {D{\"u}rig}, {Enge},
  {Fran{\c{c}}ois}, {Fourmond}, {Gillis}, {Giordanengo}, {Gissot}, {Green},
  {Guerreiro}, {Guilbaud}, {Gyo}, {Haberreiter}, {Hafiz}, {Hailey}, {Halain},
  {Hansotte}, {Hecquet}, {Heerlein}, {Hellin}, {Hemsley}, {Hermans}, {Hervier},
  {Hochedez}, {Houbrechts}, {Ihsan}, {Jacques}, {J{\'e}r{\^o}me}, {Jones},
  {Kahle}, {Kennedy}, {Klaproth}, {Kolleck}, {Koller}, {Kotsialos},
  {Kraaikamp}, {Langer}, {Lawrenson}, {Le Clech'}, {Lenaerts}, {Liebecq},
  {Linder}, {Long}, {Mampaey}, {Markiewicz-Innes}, {Marquet}, {Marsch},
  {Matthews}, {Mazy}, {Mazzoli}, {Meining}, {Meltchakov}, {Mercier}, {Meyer},
  {Monecke}, {Monfort}, {Morinaud}, {Moron}, {Mountney}, {M{\"u}ller},
  {Nicula}, {Parenti}, {Peter}, {Pfiffner}, {Philippon}, {Phillips},
  {Plesseria}, {Pylyser}, {Rabecki}, {Ravet-Krill}, {Rebellato}, {Renotte},
  {Rodriguez}, {Roose}, {Rosin}, {Rossi}, {Roth}, {Rouesnel}, {Roulliay},
  {Rousseau}, {Ruane}, {Scanlan}, {Schlatter}, {Seaton}, {Silliman}, {Smit},
  {Smith}, {Solanki}, {Spescha}, {Spencer}, {Stegen}, {Stockman}, {Szwec},
  {Tamiatto}, {Tandy}, {Teriaca}, {Theobald}, {Tychon}, {van Driel-Gesztelyi},
  {Verbeeck}, {Vial}, {Werner}, {West}, {Westwood}, {Wiegelmann}, {Willis},
  {Winter}, {Zerr}, {Zhang}, \& {Zhukov}}]{Rochus_etal:2020}
{Rochus}, P., {Auch{\`e}re}, F., {Berghmans}, D., {et~al.} 2020, \aap, 642, A8

\bibitem[{{Rouppe van der Voort} {et~al.}(2017){Rouppe van der Voort}, {De
  Pontieu}, {Scharmer}, {de la Cruz Rodr{\'\i}guez}, {Mart{\'\i}nez-Sykora},
  {N{\'o}brega-Siverio}, {Guo}, {Jafarzadeh}, {Pereira}, {Hansteen},
  {Carlsson}, \& {Vissers}}]{Rouppe-van-der-Voort_etal:2017}
{Rouppe van der Voort}, L., {De Pontieu}, B., {Scharmer}, G.~B., {et~al.} 2017,
  \apjl, 851, L6

\bibitem[{{Sand} {et~al.}(2025){Sand}, {Rouppe van der Voort}, {Joshi}, {Bose},
  {N{\'o}brega-Siverio}, \& {Gri{\~n}{\'o}n-Mar{\'\i}n}}]{Sand_etal:2025}
{Sand}, M.~O., {Rouppe van der Voort}, L. H.~M., {Joshi}, J., {et~al.} 2025,
  \aap, 697, A180

\bibitem[{{Scharmer} {et~al.}(2003){Scharmer}, {Bjelksjo}, {Korhonen},
  {Lindberg}, \& {Petterson}}]{Scharmer_etal:2003}
{Scharmer}, G.~B., {Bjelksjo}, K., {Korhonen}, T.~K., {Lindberg}, B., \&
  {Petterson}, B. 2003, in Society of Photo-Optical Instrumentation Engineers
  (SPIE) Conference Series, Vol. 4853, Innovative Telescopes and
  Instrumentation for Solar Astrophysics, ed. S.~L. {Keil} \& S.~V. {Avakyan},
  341--350

\bibitem[{{Scharmer} {et~al.}(2019){Scharmer}, {L{\"o}fdahl}, {Sliepen}, \& {de
  la Cruz Rodr{\'\i}guez}}]{Scharmer_etal:2019}
{Scharmer}, G.~B., {L{\"o}fdahl}, M.~G., {Sliepen}, G., \& {de la Cruz
  Rodr{\'\i}guez}, J. 2019, \aap, 626, A55

\bibitem[{{Scharmer} {et~al.}(2008){Scharmer}, {Narayan}, {Hillberg}, {de la
  Cruz Rodriguez}, {L{\"o}fdahl}, {Kiselman}, {S{\"u}tterlin}, {van Noort}, \&
  {Lagg}}]{Scharmer_etal:2008}
{Scharmer}, G.~B., {Narayan}, G., {Hillberg}, T., {et~al.} 2008, \apjl, 689,
  L69

\bibitem[{{Scherrer} {et~al.}(2012){Scherrer}, {Schou}, {Bush}, {Kosovichev},
  {Bogart}, {Hoeksema}, {Liu}, {Duvall}, {Zhao}, {Title}, {Schrijver},
  {Tarbell}, \& {Tomczyk}}]{Scherrer_etal:2012}
{Scherrer}, P.~H., {Schou}, J., {Bush}, R.~I., {et~al.} 2012, \solphys, 275,
  207

\bibitem[{{Schindler} {et~al.}(1988){Schindler}, {Hesse}, \&
  {Birn}}]{Schindler_etal:1988}
{Schindler}, K., {Hesse}, M., \& {Birn}, J. 1988, \jgr, 93, 5547

\bibitem[{{Schmidt} {et~al.}(2012){Schmidt}, {von der L{\"u}he}, {Volkmer},
  {Denker}, {Solanki}, {Balthasar}, {Bello Gonz{\'a}lez}, {Berkefeld},
  {Collados}, {Fischer}, {Halbgewachs}, {Heidecke}, {Hofmann}, {Kneer}, {Lagg},
  {Nicklas}, {Popow}, {Puschmann}, {Schmidt}, {Sigwarth}, {Sobotka}, {Soltau},
  {Staude}, {Strassmeier}, \& {Waldmann}}]{Schmidt_etal:2012}
{Schmidt}, W., {von der L{\"u}he}, O., {Volkmer}, R., {et~al.} 2012,
  Astronomische Nachrichten, 333, 796

\bibitem[{{Sen} \& {Moreno-Insertis}(2025)}]{Sen_Moreno-Insertis:2025}
{Sen}, S. \& {Moreno-Insertis}, F. 2025, \aap, 699, A106

\bibitem[{{Sen} {et~al.}(2026){Sen}, {Sukarmadji}, {N{\'o}brega-Siverio},
  {Moreno-Insertis}, {Mart{\'\i}nez-Sykora}, \& {Antolin}}]{Sen_etal:2026}
{Sen}, S., {Sukarmadji}, A. R.~C., {N{\'o}brega-Siverio}, D., {et~al.} 2026,
  \apjl, 997, L50

\bibitem[{{Shibata} \& {Magara}(2011)}]{Shibata_Magara:2011}
{Shibata}, K. \& {Magara}, T. 2011, Living Reviews in Solar Physics, 8, 6

\bibitem[{{Shimizu} {et~al.}(2020){Shimizu}, {Imada}, {Kawate}, {Suematsu},
  {Hara}, {Tsuzuki}, {Katsukawa}, {Kubo}, {Ishikawa}, {Watanabe}, {Toriumi},
  {Ichimoto}, {Nagata}, {Hasegawa}, {Yokoyama}, {Watanabe}, {Tsuno},
  {Korendyke}, {Warren}, {De Pontieu}, {Boerner}, {Solanki}, {Teriaca},
  {Schuehle}, {Matthews}, {Long}, {Thomas}, {Hancock}, {Reid}, {Fludra},
  {Auch{\`e}re}, {Andretta}, {Naletto}, {Poletto}, \&
  {Harra}}]{Shimizu_etal:2020}
{Shimizu}, T., {Imada}, S., {Kawate}, T., {et~al.} 2020, in Society of
  Photo-Optical Instrumentation Engineers (SPIE) Conference Series, Vol. 11444,
  Space Telescopes and Instrumentation 2020: Ultraviolet to Gamma Ray, ed.
  J.-W.~A. {den Herder}, S.~{Nikzad}, \& K.~{Nakazawa}, 114440N

\bibitem[{{Sinjan} {et~al.}(2024){Sinjan}, {Solanki}, {Hirzberger},
  {Riethm{\"u}ller}, \& {Przybylski}}]{Sinjan_etal:2024}
{Sinjan}, J., {Solanki}, S.~K., {Hirzberger}, J., {Riethm{\"u}ller}, T.~L., \&
  {Przybylski}, D. 2024, \aap, 690, A341

\bibitem[{{Skartlien} {et~al.}(2000){Skartlien}, {Stein}, \&
  {Nordlund}}]{Skartlien_etal:2000}
{Skartlien}, R., {Stein}, R.~F., \& {Nordlund}, {\r{A}}. 2000, \apj, 541, 468

\bibitem[{{Solanki} {et~al.}(2020){Solanki}, {del Toro Iniesta}, {Woch},
  {Gandorfer}, {Hirzberger}, {Alvarez-Herrero}, {Appourchaux}, {Mart{\'\i}nez
  Pillet}, {P{\'e}rez-Grande}, {Sanchis Kilders}, {Schmidt}, {G{\'o}mez Cama},
  {Michalik}, {Deutsch}, {Fernandez-Rico}, {Grauf}, {Gizon}, {Heerlein},
  {Kolleck}, {Lagg}, {Meller}, {M{\"u}ller}, {Sch{\"u}hle}, {Staub}, {Albert},
  {Alvarez Copano}, {Beckmann}, {Bischoff}, {Busse}, {Enge}, {Frahm},
  {Germerott}, {Guerrero}, {L{\"o}ptien}, {Meierdierks}, {Oberdorfer},
  {Papagiannaki}, {Ramanath}, {Schou}, {Werner}, {Yang}, {Zerr}, {Bergmann},
  {Bochmann}, {Heinrichs}, {Meyer}, {Monecke}, {M{\"u}ller}, {Sperling},
  {{\'A}lvarez Garc{\'\i}a}, {Aparicio}, {Balaguer Jim{\'e}nez}, {Bellot
  Rubio}, {Cobos Carracosa}, {Girela}, {Hern{\'a}ndez Exp{\'o}sito}, {Herranz},
  {Labrousse}, {L{\'o}pez Jim{\'e}nez}, {Orozco Su{\'a}rez}, {Ramos},
  {Barandiar{\'a}n}, {Bastide}, {Campuzano}, {Cebollero}, {D{\'a}vila},
  {Fern{\'a}ndez-Medina}, {Garc{\'\i}a Parejo}, {Garranzo-Garc{\'\i}a},
  {Laguna}, {Mart{\'\i}n}, {Navarro}, {N{\'u}{\~n}ez Peral}, {Royo},
  {S{\'a}nchez}, {Silva-L{\'o}pez}, {Vera}, {Villanueva}, {Fourmond}, {de
  Galarreta}, {Bouzit}, {Hervier}, {Le Clec'h}, {Szwec}, {Chaigneau},
  {Buttice}, {Dominguez-Tagle}, {Philippon}, {Boumier}, {Le Cocguen},
  {Baranjuk}, {Bell}, {Berkefeld}, {Baumgartner}, {Heidecke}, {Maue}, {Nakai},
  {Scheiffelen}, {Sigwarth}, {Soltau}, {Volkmer}, {Blanco Rodr{\'\i}guez},
  {Domingo}, {Ferreres Sabater}, {Gasent Blesa}, {Rodr{\'\i}guez
  Mart{\'\i}nez}, {Osorno Caudel}, {Bosch}, {Casas}, {Carmona}, {Herms},
  {Roma}, {Alonso}, {G{\'o}mez-Sanjuan}, {Piqueras}, {Torralbo}, {Fiethe},
  {Guan}, {Lange}, {Michel}, {Bonet}, {Fahmy}, {M{\"u}ller}, \&
  {Zouganelis}}]{Solanki_etal:2020}
{Solanki}, S.~K., {del Toro Iniesta}, J.~C., {Woch}, J., {et~al.} 2020, \aap,
  642, A11

\bibitem[{{Sukarmadji} {et~al.}(2022){Sukarmadji}, {Antolin}, \&
  {McLaughlin}}]{Sukarmadji_etal:2022}
{Sukarmadji}, A. R.~C., {Antolin}, P., \& {McLaughlin}, J.~A. 2022, \apj, 934,
  190

\bibitem[{{Tam} {et~al.}(2015){Tam}, {Hood}, {Browning}, \&
  {Cargill}}]{Tam_etal:2015}
{Tam}, K.~V., {Hood}, A.~W., {Browning}, P.~K., \& {Cargill}, P.~J. 2015, \aap,
  580, A122

\bibitem[{{Titov}(2007)}]{Titov:2007}
{Titov}, V.~S. 2007, \apj, 660, 863

\bibitem[{{Titov} {et~al.}(2002){Titov}, {Hornig}, \&
  {D{\'e}moulin}}]{Titov_etal:2002}
{Titov}, V.~S., {Hornig}, G., \& {D{\'e}moulin}, P. 2002, Journal of
  Geophysical Research (Space Physics), 107, 1164

\bibitem[{{Tiwari} {et~al.}(2021){Tiwari}, {Evans}, {Panesar}, {Prasad}, \&
  {Moore}}]{Tiwari_etal:2021}
{Tiwari}, S.~K., {Evans}, C.~L., {Panesar}, N.~K., {Prasad}, A., \& {Moore},
  R.~L. 2021, \apj, 908, 151

\bibitem[{{Tiwari} {et~al.}(2019){Tiwari}, {Panesar}, {Moore}, {De Pontieu},
  {Winebarger}, {Golub}, {Savage}, {Rachmeler}, {Kobayashi}, {Testa}, {Warren},
  {Brooks}, {Cirtain}, {McKenzie}, {Morton}, {Peter}, \&
  {Walsh}}]{Tiwari_etal:2019}
{Tiwari}, S.~K., {Panesar}, N.~K., {Moore}, R.~L., {et~al.} 2019, \apj, 887, 56

\bibitem[{{T{\"o}r{\"o}k} {et~al.}(2004){T{\"o}r{\"o}k}, {Kliem}, \&
  {Titov}}]{Torok_etal:2004}
{T{\"o}r{\"o}k}, T., {Kliem}, B., \& {Titov}, V.~S. 2004, \aap, 413, L27

\bibitem[{{van Noort} {et~al.}(2022){van Noort}, {Bischoff}, {Kramer},
  {Solanki}, \& {Kiselman}}]{van-Noort_etal:2022a}
{van Noort}, M., {Bischoff}, J., {Kramer}, A., {Solanki}, S.~K., \& {Kiselman},
  D. 2022, \aap, 668, A149

\bibitem[{{van Noort} \& {Chanumolu}(2022)}]{van-Noort_etal:2022b}
{van Noort}, M. \& {Chanumolu}, A. 2022, \aap, 668, A150

\bibitem[{Virtanen {et~al.}(2020)Virtanen, Gommers, Oliphant, Haberland, Reddy,
  Cournapeau, Burovski, Peterson, Weckesser, Bright, {van der Walt}, Brett,
  Wilson, Millman, Mayorov, Nelson, Jones, Kern, Larson, Carey, Polat, Feng,
  Moore, {VanderPlas}, Laxalde, Perktold, Cimrman, Henriksen, Quintero, Harris,
  Archibald, Ribeiro, Pedregosa, {van Mulbregt}, \& {SciPy 1.0
  Contributors}}]{Scipy:2020}
Virtanen, P., Gommers, R., Oliphant, T.~E., {et~al.} 2020, Nature Methods, 17,
  261

\bibitem[{{Weberg} {et~al.}(2023){Weberg}, {Warren}, {Crump}, \&
  {Barnes}}]{Weberg_etal:2023}
{Weberg}, M., {Warren}, H., {Crump}, N., \& {Barnes}, W. 2023, The Journal of
  Open Source Software, 8, 4914

\bibitem[{{Wyper} {et~al.}(2017){Wyper}, {Antiochos}, \&
  {DeVore}}]{Wyper_etal:2017}
{Wyper}, P.~F., {Antiochos}, S.~K., \& {DeVore}, C.~R. 2017, \nat, 544, 452

\bibitem[{{Wyper} \& {Hesse}(2015)}]{Wyper_Hesse:2015}
{Wyper}, P.~F. \& {Hesse}, M. 2015, Physics of Plasmas, 22, 042117

\bibitem[{{Yokoyama} \& {Shibata}(1996)}]{Yokoyama_Shibata:1996}
{Yokoyama}, T. \& {Shibata}, K. 1996, \pasj, 48, 353

\bibitem[{{Young} {et~al.}(2018){Young}, {Tian}, {Peter}, {Rutten}, {Nelson},
  {Huang}, {Schmieder}, {Vissers}, {Toriumi}, {Rouppe van der Voort},
  {Madjarska}, {Danilovic}, {Berlicki}, {Chitta}, {Cheung}, {Madsen},
  {Reardon}, {Katsukawa}, \& {Heinzel}}]{Young_etal:2018}
{Young}, P.~R., {Tian}, H., {Peter}, H., {et~al.} 2018, \ssr, 214, 120

\bibitem[{{Zhang} {et~al.}(2001){Zhang}, {Wang}, {Deng}, \&
  {Wu}}]{Zhang_etal:2001}
{Zhang}, J., {Wang}, J., {Deng}, Y., \& {Wu}, D. 2001, \apjl, 548, L99

\bibitem[{{Zhang} {et~al.}(2022){Zhang}, {Chen}, {Liu}, \&
  {Wang}}]{Zhang_etal:2022}
{Zhang}, P., {Chen}, J., {Liu}, R., \& {Wang}, C. 2022, \apj, 937, 26

\bibitem[{{Zhang} {et~al.}(2012){Zhang}, {Chen}, {Guo}, {Fang}, \&
  {Ding}}]{Zhang_etal:2012}
{Zhang}, Q.~M., {Chen}, P.~F., {Guo}, Y., {Fang}, C., \& {Ding}, M.~D. 2012,
  \apj, 746, 19

\bibitem[{{Zhu} {et~al.}(2025){Zhu}, {Harra}, {Barczynski}, {Janitzek},
  {Plowman}, {Mzerguat}, {Auch{\`e}re}, {Thompson}, {Parenti}, {Chitta},
  {Peter}, {Fredvik}, {Grundy}, {Ni}, {Chen}, \& {Valori}}]{Zhu_etal:2025}
{Zhu}, Y., {Harra}, L., {Barczynski}, K., {et~al.} 2025, \aap, 701, A205

\end{thebibliography}

\begin{appendix}
\onecolumn
\nolinenumbers 
\section{Magnetic connectivity of the quarter-note-shaped upflow}\label{app:referee}
\begin{figure*}[!ht]
   \centering
   \includegraphics[width=1.0\textwidth]{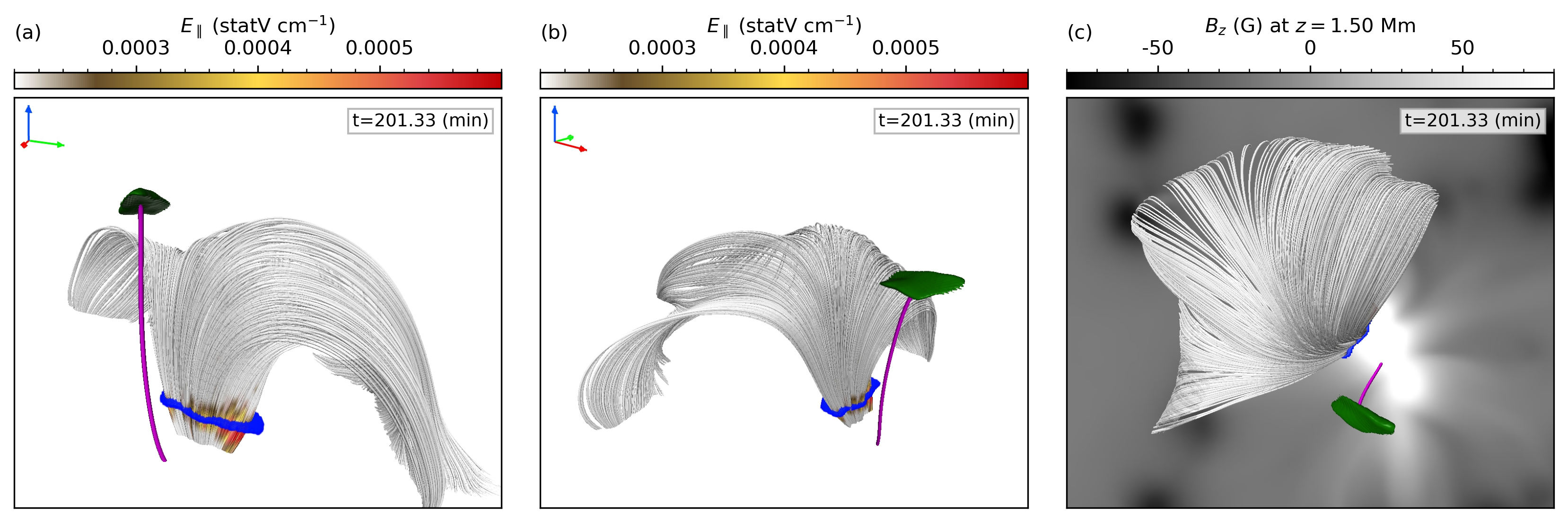}
   \caption{Magnetic connectivity of the full quarter-note-shaped upflow region at $t=201.33$~min from three different viewing angles.
   The magnetic field lines associated with the upflow are colored by $\eparb$.
   The inner spine is shown in magenta, while the green isosurface of $B=1$~G encloses the null-point region.
   The red--green--blue axes indicate the $x$-$y$-$z$ directions.
   In panel (c), the view from above is complemented by a horizontal map of $B_z$ at $z=1.5$~Mm.
   An \href{online}{online} animation shows the evolution from $t=196.67$~min to $t=201.67$~min.
   }
   \label{f:figure_appendix}
\end{figure*}

Figure~\ref{f:figure_appendix} presents a complete perspective of the magnetic field connectivity associated with the quarter-note-shaped upflow described
in Sect.~\ref{subs:results_reconnection}.
Here, a much denser set of magnetic field lines than in Fig.~\ref{f:figure_02} is traced from seed points covering the full upflow patch at $z=1.5$~Mm (shown in blue), following the same procedure as in Fig.~\ref{f:figure_02}, with the field lines integrated both upward and downward and the downward tracing stopped at $z=0.7$~Mm.
The field lines are again colored by $E_\parallel$, showing that the strongest signal remains concentrated at low heights around $z\sim1.5$~Mm.
Two additional features are included: the inner spine (magenta line) and an isosurface of $B=1$~G that encloses the null-point region.
Panels (a)--(c) show three different viewing angles; in panel (c), the view from above is complemented by a horizontal map of $B_z$ at $z=1.5$~Mm, which helps to locate the upflow region and the inner spine with respect to the parasitic polarity of the CBP.
The figure illustrates that the field lines associated with the upflow remain confined to low-lying loops and do not connect to the null-point region.

\section{Upflow identification and tracking}\label{app:tracking_method}
The animation of Fig.~\ref{f:figure_01} shows that the CBP develops recurrent 
episodes of hot upflows around the parasitic polarity, which are examined in more detail 
in Sect.~\ref{subs:results_tracking} using Fig.~\ref{f:figure_04}.
To identify and track such upflowing structures within the central region of the CBP, 
we restricted the analysis to a square region defined by $x\in[11.0,19.9]$~Mm and $y\in[13.0,21.9]$~Mm at $z=1.5$~Mm.
At each snapshot, we constructed a binary mask selecting pixels with $u_z>10$~km~s$^{-1}$.
We identified spatially connected regions using the SciPy routine \texttt{scipy.ndimage.label} \citep{Scipy:2020}, 
adopting eight-neighbor connectivity so that pixels that share either an edge or a corner are considered 
part of the same region, hereafter referred to as an island.
We retained only islands containing at least 10 pixels.
Given the horizontal grid spacing, $\Delta x=\Delta y=62.5$~km, this corresponds 
to a minimum area of approximately $0.039$~Mm$^2$.
This criterion removes very small detections and avoids placing emphasis on structures close to the 
minimum spatial scales allowed by the numerical grid.
To track the islands in time, we compared each island at time $t_i$ with those detected after a time interval $\Delta t$, 
such that $t_{i+1}=t_i+\Delta t$.
Here, we used $\Delta t=20$~s, corresponding to every second simulation snapshot.
To accommodate small displacements and changes in shape between consecutive analyzed snapshots, the footprint of each island 
at $t_i$ is expanded by one pixel in all directions using \texttt{scipy.ndimage.binary\_dilation} \citep{Scipy:2020}.
A temporal connection is established with every island at $t_{i+1}$ whose footprint overlaps the expanded region 
by at least one pixel.
The algorithm naturally accounts for both splitting and merging events.
All islands connected through successive continuations, splits, or mergers are considered part of the same upflow event.
Thus, an upflow event begins when an island has no predecessor in the previous analyzed snapshot and ends when 
none of its constituent islands has a successor in the following one.

\section{Useful quantities}\label{app:useful_quantities}
\subsection{The squashing factors \texorpdfstring{$Q$}{Q} and \texorpdfstring{$Q_{\perp}$}{Qperp}}
The squashing factor $Q$ of a vector field measures how much neighboring field lines diverge or converge as they 
connect one footpoint region to another \citep{Titov_etal:2002,Titov:2007}.
It is useful to find separatrices and QSLs \citep[][]{Priest_Demoulin:1995} and is defined as follows:

\begin{equation}
Q(x_1, y_1) = \frac{a^2 + b^2 + c^2 + d^2}{\left| \det D_{12} \right|},
\label{eq:Q}
\end{equation}
where the coefficients $a$, $b$, $c$, $d$ and the Jacobian matrix \( D_{12} \) of the field line mapping are given by
\[
D_{12} =
\begin{pmatrix}
\frac{\partial x_2}{\partial x_1} & \frac{\partial x_2}{\partial y_1} \\
\frac{\partial y_2}{\partial x_1} & \frac{\partial y_2}{\partial y_1}
\end{pmatrix}
\equiv
\begin{pmatrix}
a & b \\
c & d
\end{pmatrix}.
\]
Here, $(x_1,y_1)$ and $(x_2,y_2)$ denote the footpoint coordinates of a magnetic field line on the two boundary 
surfaces between which the mapping is computed.

In regions where the magnetic field is highly inclined with respect to the evaluation surface, such as in the 
presence of bald patches, $Q$ may be artificially enhanced due to geometrical projection effects \citep{Titov:2007}. 
In such cases, the perpendicular squashing factor $Q_{\perp}$, which provides the squashing in the directions 
perpendicular to the field lines, is a more reliable indicator since it removes the projection effect on 
boundaries \citep{Titov:2007,Zhang_etal:2022}.
Thus, in this study we used $Q_{\perp}$, computed with \texttt{FastQSL2}\footnote{\url{https://github.com/el2718/FastQSL2}}, 
a recently released and optimized code developed by \citet{Chen_etal:2026}. 
This code builds on the earlier QSL-computation tools of \citet{Liu_etal:2016} and \citet{Zhang_etal:2022}.

\subsection{Twist parameters \texorpdfstring{$t_w$}{tw} and \texorpdfstring{$\mathcal{T}_w$}{Tw}}
The twist density parameter  is defined as 
\begin{equation}\label{eq:twist_density}
    t_w =  \frac{\mu_0}{4\pi}\,\frac{\vec{J} \cdot \vec{B} }{B^2}.
\end{equation}
It has the dimension of an inverse length and provides a local measure of how much the magnetic field 
is twisted around its axis \citep{Berger_Prior_2006}. 
For a perfectly force-free field, this parameter is essentially equivalent (modulo a constant)
to the force-free $\alpha$-parameter, where $\vec{J} = \alpha \vec{B}$.

The field-line twist number $\mathcal{T}_w$ quantifies how many turns a given magnetic field line makes around 
a neighboring infinitesimally close one along its path. 
It is defined as the line integral of the twist density,
\begin{equation}
\mathcal{T}_w = \frac{\mu_0}{4\pi}\,\int_L \frac{\vec{J} \cdot \vec{B} }{B^2} \, dl =  \int_L t_w \, dl.
\end{equation}
This quantity is also provided by the \texttt{FastQSL2} code.

\end{appendix}

\end{document}